\documentclass{article}

\usepackage{arxiv}

\usepackage[utf8]{inputenc} 
\usepackage[T1]{fontenc}    
\usepackage{hyperref}       
\usepackage{url}            
\usepackage{booktabs}       
\usepackage{amsfonts}       
\usepackage{nicefrac}       
\usepackage{microtype}      
\usepackage{graphicx}
\usepackage{multirow}
\usepackage{subfig}
\usepackage{doi}
\usepackage{multicol}

\graphicspath{ {./images/} }

\title{Misrepresenting scientific consensus on COVID-19: The amplification of dissenting scientists on Twitter}

\author{Alexandros Efstratiou \\
	University College London\\
	\AND
	Tristan Caulfield \\
	University College London \\
}

\date{}

\renewcommand{\headeright}{}
\renewcommand{\undertitle}{}
\renewcommand{\shorttitle}{Creating false consensus around COVID-19 on Twitter}

\hypersetup{
pdftitle={Misrepresenting Scientific Consensus on COVID-19: The Amplification of Dissenting Scientists on Twitter},
pdfsubject={q-bio.NC, q-bio.QM},
pdfauthor={Alexandros Efstratiou, Tristan Caulfield},
pdfkeywords={Misinformation, False consensus, Amplification, Social networks},
}

\begin{document}
\maketitle

\begin{abstract}
    The COVID-19 pandemic has resulted in a slew of misinformation, often described as an ``infodemic''.
    Whereas previous research has focused on the propagation of unreliable sources as a main vehicle of misinformation, the present study focuses on exploring the role of scientists whose views oppose the scientific consensus.
    Using Nobelists in Physiology and Medicine as a proxy for scientific consensus, we analyze two separate datasets: 15.8K tweets by 13.1K unique users on COVID-19 vaccines specifically, and 208K tweets by 151K unique users on COVID-19 broadly which mention the Nobelist names.
    Our analyses reveal that dissenting scientists are amplified by a factor of 426 relative to true scientific consensus in the context of COVID-19 vaccines, and by a factor of 43 in the context of COVID-19 generally.
    Although more popular accounts tend to mention consensus-abiding scientists more, our results suggest that this false consensus is driven by higher engagement with dissent-mentioning tweets.
    Furthermore, false consensus mostly occurs due to traffic spikes following highly popularized statements of dissenting scientists.
    We find that dissenting voices are mainly discussed in French, English-speaking, Turkish, Brazilian, Argentine, Indian, and Japanese misinformation clusters.
    This research suggests that social media platforms should prioritize the exposure of consensus-abiding scientists as a vehicle of reversing false consensus and addressing misinformation stemming from seemingly credible sources.
\end{abstract}


\section{Introduction}\label{sec:intro}

As of November 2021, the COVID-19 virus has infected over 252M people and claimed over 5.1M lives.
The vaccination program against the pandemic is still ongoing, with 7.16B vaccine doses having been administered so far~\cite{noauthor_who_2021}.
However, hesitancy about these particular vaccines is alarming, with percentages of at least somewhat concerned citizens as high as 46.3\% in Italy, 43.1\% in the US, and 41.1\% in France~\cite{sallam_covid-19_2021}.
While these vaccines are being recommended by official health organizations based on scientific consensus around their safety and efficacy, there seems to be some concern even among a minority of medically literate people.
Recently, Lucia and colleagues~\cite{lucia_covid-19_2020} showed that COVID-19 vaccine hesitancy rates among a sample of medical students was at about 23\%. 

Much like is the case with climate change~\cite{bayes_research_2020}, it is likely that the minority of scientists who dissent from consensus may disproportionately be shaping public opinion on COVID-19.
Pickles et al.~\cite{pickles_covid-19_2021} have found that belief in COVID-19 misinformation is associated with distrust towards official government and scientific institutions.
Similarly, people who trust scientists and ``conventional wisdom'' about COVID-19 are less susceptible to misinformation~\cite{enders_different_2020,roozenbeek_susceptibility_2020}.
If a scientist opposes consensus, they could be perceived as less ``corrupt'' and more trustworthy by individuals who do not subscribe to official institutions.
Similarly, Salali and Uysal~\cite{salali_covid-19_2020} have reported that COVID-19 vaccine acceptance is associated with higher belief in official statements about the pandemic, for example on the origins of the virus.
Dissenting scientists could therefore be lending credibility to otherwise more conspiratorial or unsubstantiated scenarios.
Generally, people with lower educational levels seem to be more hesitant to COVID-19 vaccines, perhaps because of misconceptions around the scientific method and scientific consensus on the vaccines~\cite{malik_determinants_2020, robertson_predictors_2021}.

This line of research complements other studies which suggest that social media are the primary propagation platforms for COVID-related misinformation.
In a South Korean sample, Lee et al.~\cite{lee_associations_2020} find that the majority of exposure to COVID-19 misinformation occurs through social networking websites or messaging platforms, especially for people with lower knowledge about COVID-19 who may be more impressionable.
Consistent with this, another study reports that higher engagement with social media increases susceptibility to COVID-19 misinformation~\cite{bridgman_causes_2020}.
People who consult unverified news sources have been found to have higher COVID-19 vaccine hesitancy rates and to be reactant to calls for vaccination in the UK~\cite{murphy_psychological_2021}.
In other research~\cite{badell-grau_investigating_2020}, it is shown that performing online searches in a reactive manner to false COVID-19 information, for example looking up things such as ``chloroquine'' or ``man-made virus'', correlates with higher coronavirus deaths per capita in six countries (Spain, UK, US, Italy, Australia, and Germany).

Indeed, social media platforms do seem to be very misinformation-laden environments.
Qualitative analyses of small Tweet samples have placed the prevalence of COVID-19 misinformation on Twitter at about 25\%~\cite{kouzy_coronavirus_2020}, while larger-scale, automated analyses have identified approximately 70\% of the early information on COVID-19 to be non-credible~\cite{al-rakhami_lies_2020}.
Others have shown that COVID-19 misinformation is more prevalent than fact-checks about this misinformation in 35 countries~\cite{cha_prevalence_2021}, while unverified information on Twitter in general proliferates faster than fact-checked information~\cite{vosoughi_spread_2018}.

Evidently, we are facing a two-pronged problem.
People with lower trust in official institutions, as well as those who are less knowledgeable and educated on COVID-19, are more prone to misinformation.
At the same time, this misinformation seems to mostly proliferate on social media, where content creation is not bound by rigorous journalistic standards.
In the present study, we investigate whether the voices of dissenting scientists are disproportionately amplified on social media.
If this is truly the case, then ostensible experts in a field can lend credibility to ideas that are rejected or at least not supported by the wider scientific community.
While prior research has focused on untrustworthy news sites or conspiratorial tendencies as the main vehicles of misinformation, here we are examining another phenomenon: False consensus.
We define false consensus as the deviation of public opinion from true scientific consensus; in other words, the disproportionate amplification of scientists who do not agree with scientific consensus on Twitter.
In turn, this can inform science communication efforts around the proportionality of scientist exposure depending on their scientific views.

Specifically, we attempt to answer the following research questions:

\begin{enumerate}
    \item[\textbf{RQ1:}] On Twitter, are dissenting scientists amplified more relative to consensus-abiding scientists in the context of COVID-19 vaccines and COVID-19 more generally? What is the extent of this false consensus, if any?
    \item[\textbf{RQ2:}] What drives exposure to different scientific views on Twitter? Is exposure proportional to the popularity of the accounts that spread these views?
    \item[\textbf{RQ3:}] What are the characteristics of the retweet network that spreads these views across the Twitter platform?
\end{enumerate} 

\section{Literature review}\label{sec:lit_review}

\subsection{The role of scientific information}\label{sec:science}

Public understanding of scientific consensus is an important endeavor and a potentially effective vehicle in addressing misinformation.
For example, the sheer amount of often conflicting information about COVID-19 by either actual or ostensible experts can drive a lot of confusion among the public~\cite{lockyer_understanding_2021}.
Furthermore, Loomba et al.~\cite{loomba_measuring_2021} find that, while COVID-19 misinformation in general has the capacity of driving vaccine hesitancy, it is misinformation attributed to scientists which is most effective in doing so (e.g., ``Scientists express concerns over vaccines'').
The opposite also holds.
Correcting perceptions of scientific consensus on climate change has been found to increase belief in anthropogenic global warming~\cite{linden_scientific_2015}, while the acceptance of various scientific propositions (e.g. climate change, risk factors of different health issues) have been found to be associated with \textit{perceived} scientific consensus on the respective topics~\cite{lewandowsky_pivotal_2013}.

This body of literature reveals two key implications.
Firstly, a lot of COVID-19 misinformation may simply be due to a misconception around the number of scientists in support or in opposition of different approaches to battling the pandemic; for example the number of scientists favoring lockdowns, and those favoring a natural herd immunity approach.
This is likely to be due to both inherent confirmation biases, but also asymmetric information exposure.
Secondly, correcting these misconceptions around true scientific consensus may be an effective way of eliciting more desirable behaviors from the public, such as compliance with protective measures and vaccination, in order to protect public health.

\subsection{Cognitive biases and false consensus}\label{sec:consensus}

The notion of scientific consensus, and by extension false consensus, remains heavily understudied in the context of COVID-19.
The false consensus effect was originally reported by Ross et al.~\cite{ross_false_1977}, who found that the average person overestimates how many other people share their own convictions.
Since then, this has been demonstrated in various settings.
Examples include climate change deniers grossly overestimating the true percentage of other deniers in the population~\cite{leviston_your_2013}, and US voters overestimating the percentage of Americans who share their opinion on whether the 2020 US presidential election was rigged~\cite{weinschenk_democratic_2021}.

With the advent of social media and online communications more generally, false consensus may now be more prevalent.
This is because some users on social media tend to become enclosed in ``echo chambers'' where the same opinions are reinforced~\cite{terren_echo_2021}.
In these cases, false consensus may become more pronounced as individuals think their echo chambers are representative of the entire information ecosystem~\cite{wojcieszak_false_2008,yeager_moderation_2019}.

There is also evidence to suggest that false consensus can be caused by users' sharing tendencies.
Gallagher et al.~\cite{gallacher_mutual_2021} find that when it comes to COVID-19, it is mostly public figures that receive the biggest exposure relative to their Twitter activity.
Medical professionals, on the other hand, are not particularly prominent in terms of exposure.
Crucially, Twitter users tend to amplify figures which are either ideologically or demographically similar to them, perhaps driving a false conception of how prominent certain opinions are.
It has also been found that those who are most prone to false consensus also tend to believe that mainstream media are hostile against their opinions~\cite{schulz_we_2020}.
In this case it is possible that, in an attempt to balance out this ostensibly biased reporting of mainstream news, these people disproportionately amplify dissenting voices to bring the information ecosystem closer to their subjectively perceived consensus.

Perceived consensus, and any disconnect between it and true consensus, can have significant effects on the pandemic response.
We have already explained how false consensus can lend credence to ``scientific'' opinion which is otherwise unsupported by science, however it can also operate on the social level.
If a misguided group of people mistakenly believes that their opinion is scientifically backed, then it should be easier to establish health norms among this group.
These perceived group norms, especially in more homogeneous groups, can drive health and COVID-related behaviors in a consistent direction~\cite{bavel_using_2020,goldberg_social_2020}.

However, we do note research which has shown that it is individual beliefs, not perceived community-wide norms that drive compliance intentions with COVID-19 measures~\cite{lees_intentions_2020}.
Therefore, it is important that these norms are perceived about common-identity groups rather than broader, generic groups.

\subsection{Previous work on COVID-19 misinformation}\label{sec:previously}

COVID-19 and the ensuing infodemic have attracted academic attention, however most studies focus on either misinformation of a conspiratorial nature, or automatic detection of misinformation.

Shahi et al.~\cite{shahi_exploratory_2021} analyze and classify misinformation based on fact-checked claims, then reverse-search tweets which contain URLs with these fact-checked claims.
They perform exploratory analyses such as the emotional profile or the most-used emojis in COVID-19 fact-checked misinformation tweets.
Al-Rakhami and Al-Amri~\cite{al-rakhami_lies_2020} build a highly accurate classifier through feature extraction using a dataset of early COVID-19 information, between January and April 2020.
Their classifier, based on user-level and tweet-level features, classifies 70.22\% of early information as non-credible.

Singh et al.~\cite{singh_understanding_2020} map out the COVID-19 information landscape using a URL-based approach where low-quality or high-quality information is labeled based on the URL domain (i.e. high-quality, low-quality, or mainstream news providers). They find that misinformation does exist on Twitter, however mainstream news sources constitute a much bigger proportion. Mainstream news are also connected to these low-quality and high-quality (i.e. health news) clusters, therefore suggesting that they can be an avenue of exposure to either better- or worse-quality information.

On conspiratorial information, Shahsavari et al.~\cite{shahsavari_conspiracy_2020} analyze the emergence of new COVID-19 conspiracy theories and plot a graph of their main communities on 4chan and Reddit.
Similarly, Havey~\cite{havey_partisan_2020} analyzes a small sample of Tweets and examines how political ideology can impact belief in COVID-19 conspiracy theories.
Memon and Carey~\cite{memon_characterizing_2020} follow a keyword-based approach comprising predominantly misinformation-laden terms and find that a significant proportion of Twitter posts comes from bots, especially posts labeled as misinformation.
Yang et al.~\cite{yang_prevalence_2020} also find that bots more commonly spread misinformation than trustworthy sites, although the vast majority of misinformation still comes from organic users.

\section{Present study}\label{sec:study}

Common to all of the above approaches is the origin of COVID-related misinformation from unverified or conspiratorial sources.
While misinformation is, by default, false and unreliable, the concept of what constitutes a verified or trustworthy source may differ between communities.
For example, a scientist would generally be associated with accuracy and trustworthiness, however their claims could still be false if they are not backed up by commensurate evidence.
Therefore, while many of these approaches are based on source as an indicator of trustworthiness (e.g. URL domain, account posting the information, information appearing on fact-checking websites etc.), there may be paradoxical situations where an otherwise credible individual spreads false information.

In this study, we probe this potential phenomenon and examine if public opinion deviates from scientific consensus on COVID-19 for the following reasons.
Firstly, is a unique opportunity to observe whether certain figures can achieve popularity simply by virtue of being dissenting voices.
Secondly, it can highlight whether ``naive science'', where the scientific credentials of the source can trump the need for substantiating their claims, is weaponized.
Thirdly, and perhaps most importantly, it can highlight a need to move away from considering misinformation purely in conspiratorial terms.
Therefore, this could point towards the need to more accurately reflect scientific consensus on social media to prevent the asymmetric amplification of dissenting voices.

\subsection{Deriving a proxy for scientific consensus}

There are two main challenges with deriving a scientific consensus on COVID-19 issues.
First, due to the sheer number of people working on the pandemic, true scientific consensus is difficult to pinpoint without prior research on the topic (in contrast, lots of data are available on things like climate change consensus; see~\cite{allen_climate_2014, cook_quantifying_2013}).
Second, scientists working on COVID-19 span multiple disciplines (e.g. public policy, epidemiology, virology, social psychology, etc.), with some being considered more prestigious or accomplished than others.
These are factors that an individual could consider when deciding whose opinion they should consult.
We note that Alwan et al.~\cite{alwan_scientific_2020} did make a call to establish consensus against natural herd immunity approaches, however their list of signatories was quite small (80 people) relative to the entire community of COVID-19, with signatories spanning multiple fields.

To address both problems of relative size and comparability of opinion, we opt for a proxy approach. We restrict the eligible scientists to form part of our ratio as those who have won a Nobel prize in Physiology or Medicine from 2000 onward.
This leaves us with a manageable number of scientists whose opinions we can qualitatively assess.
Moreover, we reason that due to the prestige associated with the Nobel prize, these scientists will be invoked mainly for their laureate status.
Therefore, while there is still quite a bit of variety in the specific disciplines that Physiology and Medicine Nobel laureates work in, we argue that the gravity of their opinions should be comparable among the general public.

\subsection{Conducting the search}

First, we listed the names of Nobel laureates in Physiology or Medicine who won the award from 2000 until 2020, as they appear on the official Nobel prize website\footnote{\href{https://www.nobelprize.org/prizes/lists/all-nobel-laureates-in-physiology-or-medicine/}{https://www.nobelprize.org/prizes/lists/all-nobel-laureates-in-physiology-or-medicine/}}.
Out of 53 eligible laureates, we found that 44 were still alive (i.e., no death date listed on their Nobel prize profiles).
For these laureates, we performed a Google search to find instances where they had gone on record expressing an opinion on COVID-19 vaccines, looking at the first three pages of results.
We examined COVID-19 vaccines specifically to maximize the comparability of opinions between the laureates and to have a more exact frame of reference.

To examine the opinion of each eligible laureate on COVID-19 vaccines we used the following search approach.
Note that, while we removed the ``vaccine'' keyword if no results of interest were returned to expand the search, we continued searching the results for COVID-19 vaccine opinions \textit{specifically}.
Keywords in OR statements were used sequentially; that is, if we received no relevant results with the combination of the keywords ``COVID-19'' and ``vaccine'', we performed another search with the keywords ``coronavirus'' and ``vaccine'', then ``COVID-19'' and ``vaccines'', and so on.

{\small
\begin{verbatim}
    1. ``(Laureate name)''
        AND
        (COVID-19 OR coronavirus)
        AND
        (vaccine OR vaccines)
            IF relevant results:
                categorize laureate
            ELIF no relevant results:
    2. ``(Laureate name)''
        AND
        (COVID-19 OR coronavirus)
\end{verbatim} }

We examined all relevant results (i.e., those including at least 2 out of the 3 search terms in the title) regardless of medium or format (i.e., including videos, blog posts, social media posts, etc).
For laureates whose opinion we could not find on Google, we also searched the platform DuckDuckGo.
We did this because of Google's PageRank algorithm which prioritizes indexing based on the trustworthiness and popularity of a news page.
We reasoned that any consensus-dissenting scientists may be given more exposure on unreliable news sites, and therefore their statements could be pushed further down.
To protect from potential bias against consensus-dissenting scientists as a result of the search platform we used (i.e. to avoid under-counting dissenting scientists), we repeated the search query procedure on this platform which employs a different search engine optimization.

\subsection{Scientific consensus ratio}

Out of the 44 Nobel laureates that we searched, we were able to find 25 who had gone on record expressing an opinion on COVID-19 vaccines. Out of those, 24 were pro-vaccines, and just one was against. The one laureate who had spoken against vaccines had gone on record multiple times, stating, among other things, that vaccines cannot be considered safe if long-term side-effects are unknown, and that the vaccines may cause antibody-dependent enhancement (a phenomenon whereby vaccines can inadvertently \textit{increase} the risk and severity of infection).

Examples of pro-vaccine statements included support for the vaccination program as the only way of ending the pandemic, calls to put faith in the scientists and institutions who developed the vaccines, reassurances that the vaccines are safe because their mechanisms are known and understood, signing on open letters to expand access to vaccines to more impoverished countries, and public appearances of laureates taking the vaccines themselves. The sources consulted in this annotation procedure, as well as the names of the laureates we were able to identify, are found in Appendix \ref{append_a}.

Using this approach, we were able to derive a proxy for a ratio of scientific consensus as $1:24$ for anti-vaccine to pro-vaccine scientists.
We then annotated tweets as anti if they mentioned the one laureate who appeared to be against vaccines, and as pro if they mentioned any of the other 24.
To ensure that these tweets are not negative towards the laureates they discuss, we performed a ground-truth validation of our approach in section~\ref{sec:main_analysis}.
Based on these, we calculated public consensus $c_p$ as $c_p = \frac{anti_{total}}{ anti_{total} + pro_{total}}$.

To calculate the factor of false consensus, $c_f$, we used the following formula,
\begin{equation}\label{eq:1}
    c_f = {\frac{c_p}{c_t}}
\end{equation}
where $c_p$ is the public (Twitter-derived) ratio and $c_t = \frac{1}{25}$.
The closer $c_f$ is to 1, the more public consensus resembles scientific consensus.

\section{Data}

We utilized a keyword-based query in Twitter's full-archive API to collect three separate datasets.

\subsection{Vaccine dataset}

The first dataset regarded COVID-19 vaccine Tweets which mentioned at least one of the 25 laureates that we identified in \S\ref{sec:study}.
Our search query consisted of three key segments: Coronavirus synonyms, vaccine synonyms, and the laureate names.
The full query we used is seen in Appendix \ref{append_b}.

We collected data between January 1st 2020 and July 5th 2021.
In total, we collected 15.8K tweets from 13.1K unique users mentioning at least one of the laureates in the context of COVID-19 vaccines.
12.7K of these tweets were retweets.

\subsection{Generic COVID-19 dataset}

Given the small size of the vaccine-specific dataset, we performed a second query after removing the vaccine synonyms to broaden our search.
While our laureate annotations were about COVID-19 vaccines specifically, we conducted another cursory search on the laureates' general opinions.
We were able to find that the anti-vaccine laureate had also popularized unverified information around the virus' origins earlier in the pandemic, while none of the other laureates had gone on record opposing consensus on COVID-19 generally before\footnote{The only exception was one of the pro-vaccine laureates who had a quote falsely claiming the virus was created in a laboratory misattributed to them. The laureate released a statement where they denied having said this.}.
Therefore, we reasoned that the broader COVID-19 was a comparable topic to COVID-19 vaccines specifically in this instance, and that the potentially bigger sample would provide more context on the Twitter discussion surrounding scientific discourse.

To maximize comparability, we also performed this search from January 1st 2020 until July 5th 2021.
We were able to collect 208K tweets which mentioned at least one of the 25 laureates in the context of COVID-19 from 151K unique users. Out of these, 172.8K were retweets.

\subsection{Baseline dataset}

For comparison purposes, we also collected baseline data prior to COVID-19 becoming a prominent discussion topic on social media.
For this purpose, we removed both COVID-19 and vaccine-related terms from our search, keeping only the laureate names.
The baseline dataset included data from May 1st 2018 to December 31st 2019. This was a comparable length of time to the other two datasets.
The end date was before the first recorded case of COVID-19 outside of China when coronavirus became a worldwide topic of discussion, and right before we began data collection for the other two datasets.
In total, we collected 196K baseline tweets by 136.9K unique authors, 159K of which were retweets. We summarize all three datasets in Table \ref{tab:data}.

\begin{table}[t!]
    \centering
    \small
    \begin{tabular}{lcclc}
    \toprule
        \textbf{Dataset} & \textbf{Tweets} & \textbf{Unique authors} & \textbf{\% retweets} & \textbf{Data collection period} \\
        \midrule
        Vaccines & 15.8K & 13.1K & 80\% & 01/01/2020 - 05/07/2020 \\
        COVID-19 general & 208K & 151K & 83\% & 01/01/2020 - 05/07/2020 \\
        Baseline & 196K & 136.9K & 81\% & 01/05/2018 - 31/12/2020 \\
        \bottomrule
        \end{tabular}
        \caption{Summary of datasets}
        \label{tab:data}
\end{table}

\paragraph{Ethical considerations.} Use of Twitter API's full archive search for this project was approved by the official Twitter platform through a Twitter developer application. This project received ethical approval from the UCL ethics committee, REC no. Z6364106/2021/07/45. Data processing is performed in accordance with GDPR provisions~\cite{noauthor_general_nodate}.

\section{False consensus analysis}\label{sec:main_analysis}

In this section, we attempt to answer RQ1 (to what extent are dissenting scientists disproportionately amplified?) and RQ2 (is exposure to dissenting scientists proportional to the popularity of the accounts which share their opinions?).

\subsection{Tweet annotation}

First, we annotated the tweets in all datasets using the Nobel laureate names that appeared in them.
We followed a simple heuristic rule to do this.
If a pro-vaccine (pro-consensus) laureate name appeared in the tweet, then it would be labeled as ``pro''.
If the anti-vaccine (anti-consensus) laureate's name appeared, then the tweet would be labeled as ``anti''.
Because of how the Twitter API handles responses some of the tweets returned did not include the laureate names in their text, but rather in associated meta-data (e.g. URL link titles, embedded video titles, etc).
Therefore, if our program could not detect a laureate name in the main text of the tweet, it would also check if the tweet included a URL, and if so, check the URL string for the Nobelist names.
The specific procedure was as follows:

\vbox{%
\begin{multicols}{2}
{ \small
\begin{verbatim}
Check tweet text
Count anti-laureate instances
IF count > 0:
    return ``anti''
    STOP
Count pro-laureate instances
IF count > 0:
        return ``pro''
        STOP
ELSE:
    Check if URL exists
    IF URL does not exist:
        return ``undefined''
    ELIF URL exists:
        Check URL string
        Count anti-laureate instances
        IF count > 0:
            return ``anti''
            STOP
        Count pro-laureate instances
            IF count > 0:
                return  ``pro''
                STOP
            ELSE:
                return ``undefined''
\end{verbatim} } 
\end{multicols} }

Throughout this process, we iteratively checked any tweets which were classified as ``undefined'' for alternative spellings of laureate names and updated the annotation strings accordingly until we reached saturation.

\paragraph{Ground-truth testing.}Before proceeding with any further analyses, we constructed a validation set for ground-truth testing from the generic COVID-19 dataset, which was broader and therefore more representative.
First, we filtered the data to only include original (i.e. no retweets, quotes, or replies), English-language tweets which had not been labeled as ``undefined'' during the automated annotation process.
From this subset, we randomly sampled 1000 tweets for manual annotation.
Annotation was performed by one researcher familiar with the laureates' general views\footnote{We recognize that this may introduce some bias. The trade-off is between potentially biased annotations and expertise with respect to recognizing which scientists received exposure through these tweets, which we attempted to incorporate in our annotation procedure.}.
The annotation procedure followed a two-step priority method.
First, we checked if the tweets promoted consensus-dissenting information (e.g., human-made virus theories, anti-vaccine sentiments, etc).
If consensus-dissenting information was advocated in the tweet, we labelled it as ``anti''.
If it remained unclear whether the tweet was pro- or anti-scientific consensus, we treated it as neutral in that respect.
In that case, we would then check if the tweet mentioned one of the 25 laureates in our list.
If it did, we would annotate it as ``pro'' or ``anti'' accordingly.
Our reasoning here was that the tweet would be giving exposure to the person it mentioned, and thus either supporting or rejecting scientific consensus, respectively.

Using this annotation process, we were able to achieve an F1 score of 0.88 between our automated and ground-truth procedures.
As this was deemed satisfactory, we did not proceed with a calibration for our automated annotation.
The confusion matrix is shown in Table \ref{tab:confusion}.
Generally, we found that the heuristic rule performed well in classifying consensus-dissenting and consensus-abiding tweets.
Examples of instances where the heuristic rule failed were posts fact-checking claims by the anti-consensus laureate (thus supporting scientific consensus despite the anti-laureate's name appearing in the tweet), and posts which mistakenly attributed misleading information to pro-consensus laureates (thus going against scientific consensus despite the pro-laureate's name appearing in the tweet).

\begin{table}[t!]
    \centering
    \small
    \begin{tabular}{ccc}
    \toprule
         & \textbf{$pro_{manual}$} & \textbf{$anti_{manual}$} \\
         \midrule
         \textbf{$pro_{auto}$} & 450 & 56 \\
         \hfill \\
         \textbf{$anti_{auto}$} & 61 & 421 \\
         \bottomrule
        \end{tabular}
        \captionsetup{width=.75\linewidth}
        \caption{Confusion matrix for ground truth annotation. We manually labeled 12 tweets as ``undefined'' and therefore removed them from the F1 analysis.}
        \label{tab:confusion}
\end{table}

\subsection{True vs. public consensus deviation}

Next, we performed a simple frequency analysis of pro and anti tweets in all three of our datasets.
To calculate a factor of false consensus, we used Equation \ref{eq:1}, where $c_p$ was derived from the frequencies (absolute numbers) of pro and anti tweets accordingly.
The actual number of tweets, along with the expected number of tweets based on $c_t$ and the deviation factor between actual and expected numbers (i.e. false consensus, $c_f$) is seen in Table \ref{tab:deviation}.

\begin{table}[!t]
    \centering
    \small
    \begin{tabular}{lllrrr}
        \toprule
        \textbf{Dataset} & \textbf{Expected anti} & \textbf{Actual anti} & \textbf{Expected pro} & \textbf{Actual pro} & \textbf{$c_f$ factor} \\
        \midrule
        Vaccines & 590.32 & 13971 & 14167.68 & 787 & 426.05 \\
        General & 7499.12 & 120193 &  179978.88 & 67285 & 42.9 \\
        Baseline & 6626.08 & 6192 & 159025.92 & 159460 & 0.93 \\
        \bottomrule
        \end{tabular}
        \caption{False consensus factor, $c_f$, calculated from expected frequencies based on $c_t$ and actual frequencies $c_p$. Sum of actual pro and actual anti tweets is lower than the number of total tweets in Table \ref{tab:data} due to instances of ``undefined'' tweets.}
        \label{tab:deviation}
\end{table}

As seen in the table, the baseline dataset was generally in line with the scientific consensus ratio $c_t$, with the expected and actual values being quite close ($c_f = 0.93$).
Since the $c_f$ value for baseline was very close to 1, we can say that public discourse on Twitter prior to the COVID-19 pandemic was ``consensus-abiding''.
In other words, any deviations from 1 in the $c_f$ of the other two datasets is unlikely to be due to prior differences in the popularity of the laureates.

With respect to the vaccine dataset, we can see a very large disproportionality such that expected and observed values are almost reversed; that is, the actual anti values approximate the expected pro values, and the expected anti values are close to the actual pro values.
This results in a very large $c_f$, such that the anti-consensus laureate is amplified by a factor of 426.05.
A similar---albeit smaller-scale---pattern is observed with regards to the general COVID-19 dataset.
In this broader context, the anti-consensus laureate is amplified by 42.9 times more relative to the consensus-abiding laureates.
Our analysis demonstrates that public consensus deviates from scientific consensus to a very large extent in both the context of COVID-19 vaccines and COVID-19 generally, especially for the former.

\subsection{Time series analysis}

\begin{figure}[!tbp]
    \centering
    \subfloat[Absolute comparison.]{\includegraphics[width=0.7\textwidth]{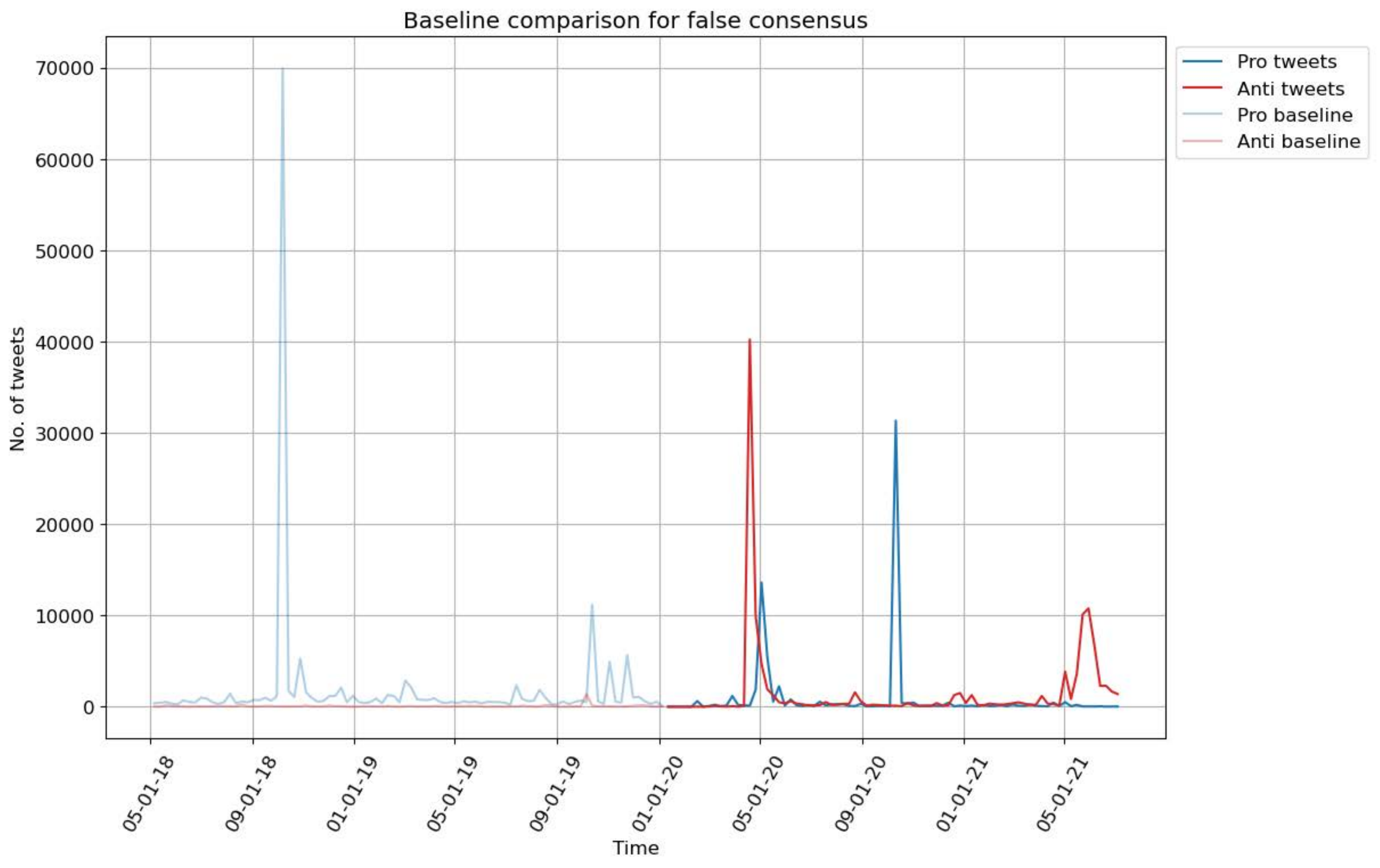}\label{fig:abscomp}}
    \hfill
    \subfloat[Log comparison.]{\includegraphics[width=0.7\textwidth]{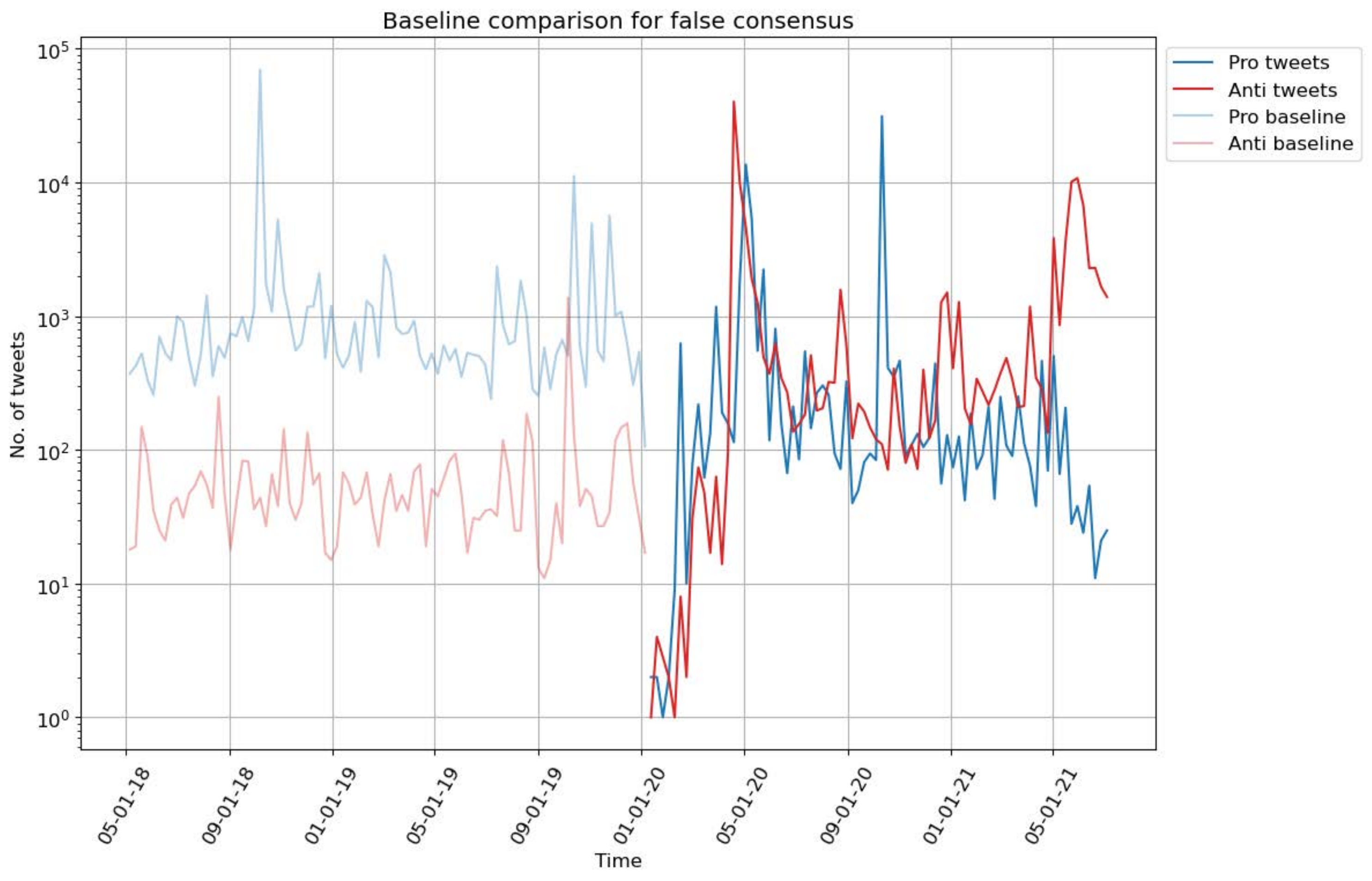}\label{fig:logcomp}}
    \caption{Frequency comparisons between baseline and general COVID-19 datasets.}
    \label{fig:comparison}
\end{figure}

To get a better idea of the dynamics regarding when these scientists are discussed on Twitter, we created time-series plots at a weekly granularity.
We first plot absolute (Figure \ref{fig:abscomp}) and log-transformed (Figure \ref{fig:logcomp}) frequency time-series comparisons between the baseline and COVID-19 general datasets.
We omit the vaccine dataset from this comparison due to its much smaller, and thus non-comparable size.

Looking at Figure \ref{fig:abscomp}, we can see that for all years included in the datasets there are spikes in pro tweets around the month of October (2018, 2019, 2020).
From a cursory examination of our data, we find that these spikes are due to the official Nobel prize announcements for the respective year which tend to attract a lot of attention and are widely retweeted.
We also observe a mild spike in pro tweets around May 2020.
This concerned a story circulating on social media which falsely claimed that one of the pro-consensus laureates said that the virus was manufactured.
On April 27th 2020, this laureate released a statement denying having said this\footnote{\href{https://www.kyoto-u.ac.jp/en/news/2020-04-27}{https://www.kyoto-u.ac.jp/en/news/2020-04-27}}.
The spike which followed was mainly due to the original story being picked up and debunked by fact-checkers, which was subsequently shared widely on Twitter.

Regarding anti tweets, we again observe that they follow a spike-like pattern.
In this case, both of these spikes followed interviews with the anti-consensus laureate which received a lot of attention.
The first spike, around mid-April 2020, concerns a public appearance of the anti-consensus laureate on CNEWS, a popular French news channel which has been said to promote conservative and far-right views.
In this appearance, dated April 17th 2020, the laureate claimed that the virus was man-made and the result of ``meticulous work'' of molecular biologists.

The second spike, occurring around late May - early June of 2021, followed an interview of the same scientist in the French documentary ``Hold-Up'', a conspiratorial film which made numerous false claims about COVID-19.
This interview was released on May 13th 2021\footnote{\href{https://planetes360.fr/pr-luc-montagnier-les-variants-viennent-des-vaccinations/}{https://planetes360.fr/pr-luc-montagnier-les-variants-viennent-des-vaccinations/}}, and it was later given to the RAIR (Rise Align Ignite Reclaim) foundation for an exclusive English translation, published on May 18th 2021\footnote{\href{https://rairfoundation.com/bombshell-nobel-prize-winner-reveals-covid-vaccine-is-creating-variants/}{https://rairfoundation.com/bombshell-nobel-prize-winner-reveals-covid-vaccine-is-creating-variants/}}.
The RAIR foundation describes itself as a grassroots activist organization and an ``integrated media platform amplifying the voices of the silent majority''.
It is a USA-based alternative media platform which promotes mainly misinformation-laden, alt-right, and anti-Muslim content.
In this interview, the laureate claimed, among other things, that vaccines are causing new coronavirus variants and that the vaccination effort is an ``unacceptable mistake''. This translation piece was the main driver of the second spike.

From Figure \ref{fig:logcomp}, which shows the number of tweets on a logarithmic scale, we can observe that prior to COVID-19 the anti-consensus laureate mentions never exceeded the total mentions for the pro-consensus laureates.
After the sharp drop in both pro-consensus and anti-consensus laureate mentions due to COVID-19 not yet having picked up traction on Twitter by January 2020, there is a rapid rise in anti tweets which is generally retained until the final date for which data is available.
The only notable exception is the Nobel prize announcement of new laureates in October 2020, when the (newly announced) pro-laureate mentions spike and briefly overtake anti-laureate mentions.

\begin{figure}[!tbp]
    \centering
    \subfloat[Absolute values.]{\includegraphics[width=0.5\textwidth]{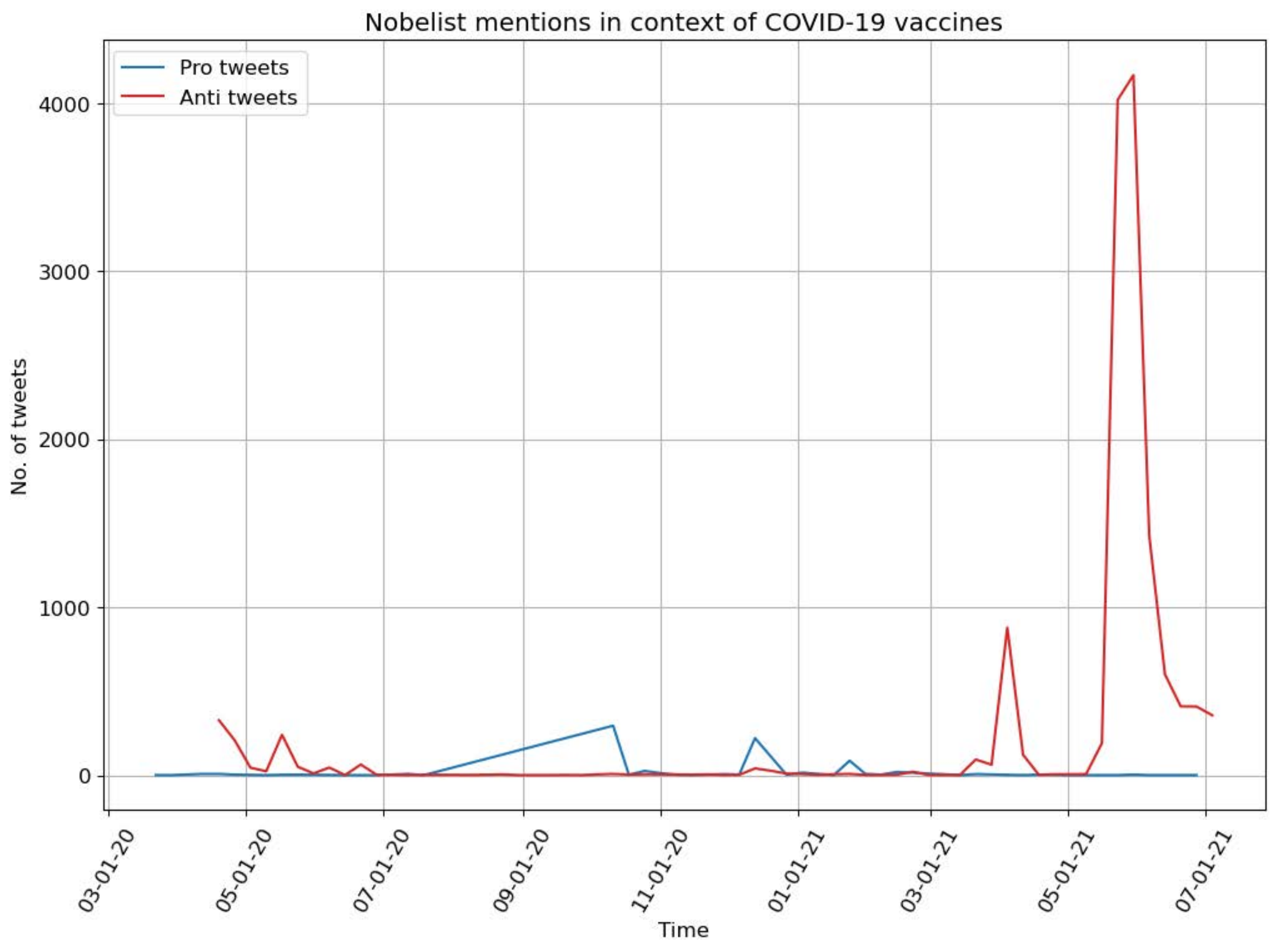}\label{fig:absvac}}
    \hfill
    \subfloat[Log-transformed values.]{\includegraphics[width=0.5\textwidth]{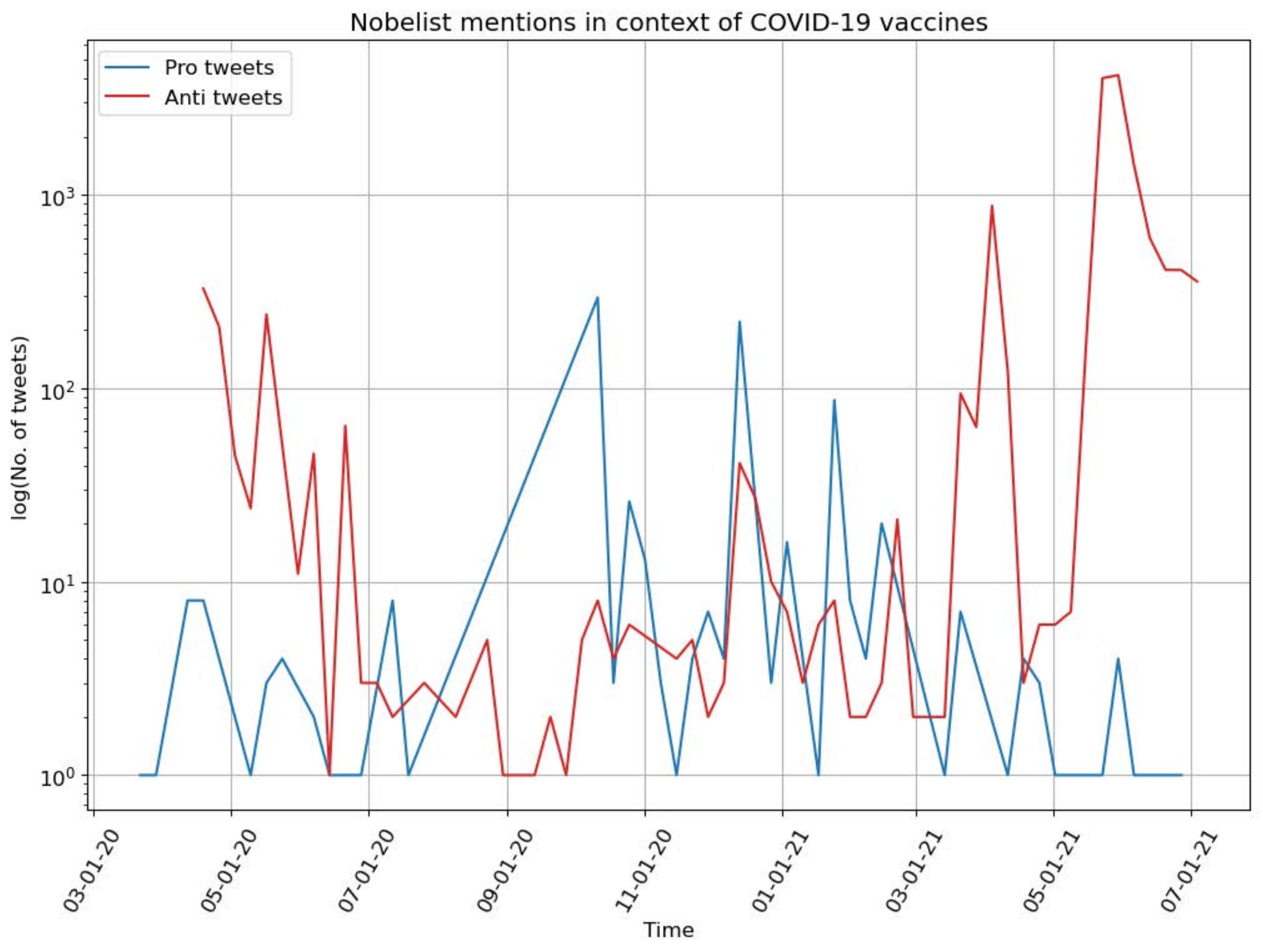}\label{fig:logvac}}
    \caption{Frequency plots for vaccine dataset.}
    \label{fig:vax}
\end{figure}

In Figure \ref{fig:vax} we also plot time-series frequency data for the vaccine dataset. The pattern is comparable with what is seen in the general COVID-19 data (Figure \ref{fig:comparison}), since we also see more attention drawn to the anti-consensus laureate around April 2020, then attention shifting to pro-consensus laureates around October 2020, and then another shift to the anti-consensus laureate following the RAIR article (Figure \ref{fig:logvac}).
However, as expected for the vaccine data, by far the most posts concern the anti-vaccine laureate's statements about the supposed dangers of the vaccination program in May 2021 (Figure \ref{fig:absvac}).

\section{Popularity metrics}\label{sec:popularity}

Our second research question concerned the drivers behind exposure to different scientists on Twitter; in other words, whether exposure is driven by the popularity of the accounts which amplify them, or if it is driven by the sheer number of users who amplify those views.

\subsection{Top posters}

\begin{figure}[!tbp]
    \centering
    \subfloat[Anti posters.]{\includegraphics[width=0.33\textwidth]{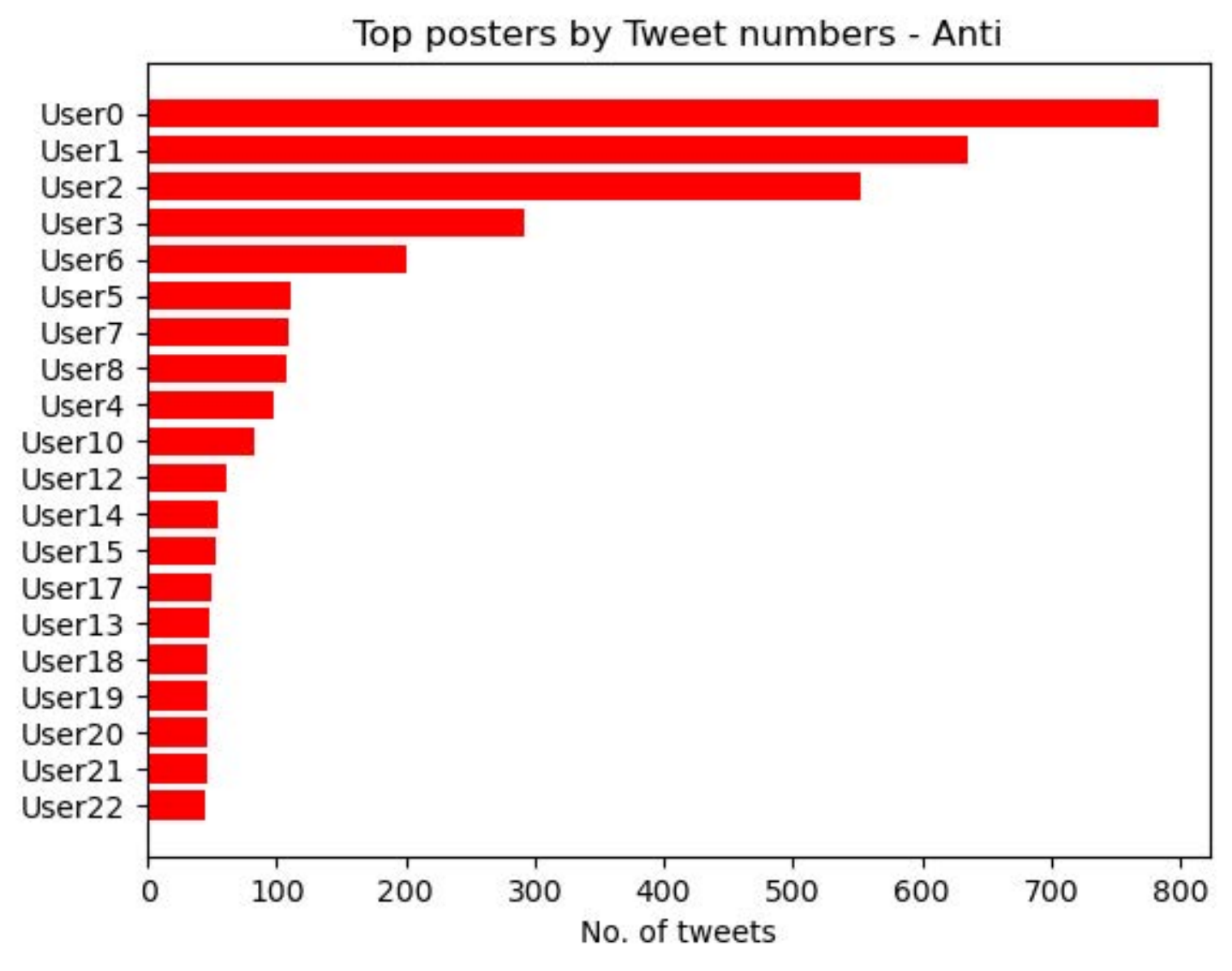}\label{fig:antiposts}}
    \hfill
    \subfloat[Pro posters.]{\includegraphics[width=0.33\textwidth]{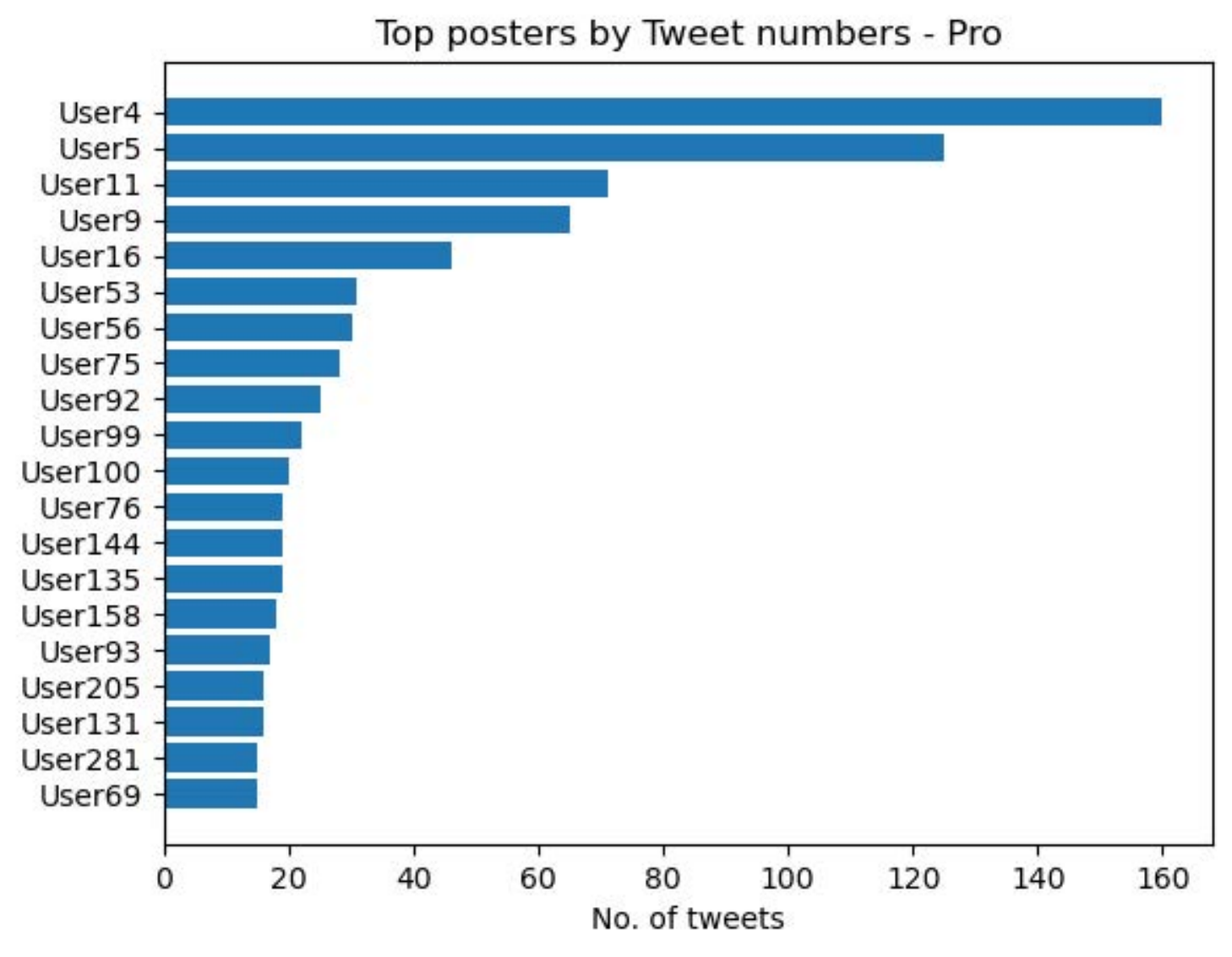}\label{fig:proposts}}
    \hfill
    \subfloat[Relative posters.]{\includegraphics[width=0.33\textwidth]{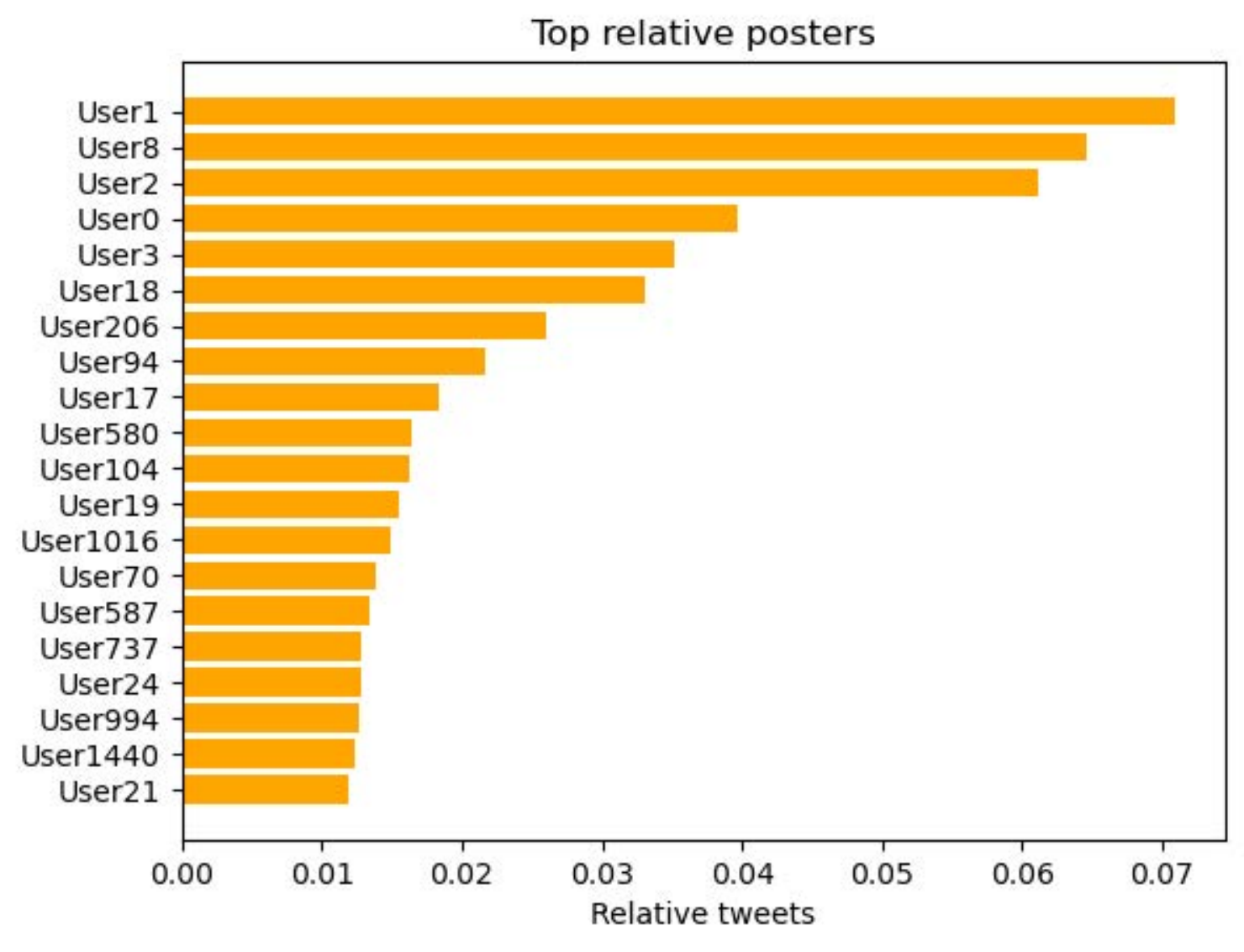}\label{fig:relposts}}
    \caption{Post frequencies of top 20 users per category.}
    \label{fig:usersposts}
\end{figure}

To get a clearer picture of how these views are disseminated on Twitter, we first (anonymously) plot the top 20 users in terms of the number of anti-consensus or pro-consensus tweets (and retweets) they produce, as well as the top users in terms of their relative number of posts (Figure \ref{fig:usersposts}).
To calculate the relative number of posts, we simply count the number of tweets in our sample produced by each user and divide them by the user's total number of tweets.
We restrict this calculation to users who have posted at least 500 total tweets in order to preserve variation in the data.
This allows for an assessment of whether the absolute number of posts is due to a general tendency to post more, or if some accounts disproportionately amplify these Nobelists in the context of COVID-19.
Because of the small sample in the COVID-19 vaccine dataset, we only show the general COVID-19 plots here.
For context, the top anti user in terms of posting frequency for COVID-19 vaccines had 47 posts, and the top pro user had just 7.

Users are indexed in descending order of the number of times they appear in our dataset, meaning User0 is the top overall poster, User1 the second-top, and so on.
As is evident from the index numbers in Figure~\ref{fig:antiposts}, the anti-consensus top posters are almost exclusively also the overall top posters.
Furthermore, the top 20 range is much higher for anti posters (approx. 50-800) than pro posters (approx. 20-160).
The top 20 anti posters post an average of 173.45 tweets, approximately 4.5 more than the top 20 pro (average 38.85 tweets).
Notice that Users 4 and 5 appear in both the anti and pro top posters.
These are news aggregation accounts which collate and share news to their audiences from across the platform in a fairly non-selective fashion.

Regarding the top 5 relative posters, we observed a suspicious pattern in the accounts' naming conventions which suggested they may be part of a botnet.
Four of these accounts are the most frequent anti-consensus posters.
We perform bot detection on these 5 accounts using the Botometer tool~\cite{davis_botornot_2016} which uses several criteria, such as the account's posting and deletion behavior, whether the account has been labeled as a spambot in other datasets, whether the account has been purchased as a follower bot, etc.
Indeed, Botometer assigns scores of 4.1, 3.8, 4.1, 4.2, and 3.8 out of 5 to these accounts respectively, pointing to high bot-like activity.
Average bot-like activity for the 20 most active accounts is 2.5 for the relative posters, 3.05 for the anti posters, and 2.88 for the pro posters.
These are relatively high figures, however this is expected as the accounts with the most posts are also more likely to be the ones that automatically aggregate and share content on Twitter.

\subsection{Post reach}

To assess post reach, we first examine whether there are any differences in the popularity of the accounts which share pro-consensus and anti-consensus tweets.
We do this for both the general and vaccine-specific datasets, and we report results both with retweets included and retweets removed.
We report the number of accounts sharing each type of tweet, along with their mean number of followers, in Table \ref{tab:followers}.
In case that an account has shared both pro- and anti-consensus tweets, it is counted in both the pro and anti account calculations.
Generally, we find that the pro-consensus accounts tend to be much more popular (i.e. have more followers), but are much fewer in numbers.
Because of the small number of accounts, and especially unique pro accounts in the vaccine-specific dataset, we only analyze the general COVID-19 dataset in the remainder of this section.

\begin{table}[t]
    \centering
    \small
    \begin{tabular}{llcccc}
    \toprule
        \multicolumn{2}{c}{\textbf{Dataset}} & \multicolumn{2}{c}{\textbf{With retweets}} & \multicolumn{2}{c}{\textbf{No retweets}} \\
         & & Followers & \textit{N} & Followers & \textit{N} \\
        \midrule
        Vaccine & Anti & 5582.11 & 11559 & 23662.98 & 2025 \\
         & Pro & 10802.20 & 760 & 56476.43 & 89 \\
        \hline
        General & Anti & 4003.49 & 83178 & 14679.94 & 15367 \\
         & Pro & 14767.46 & 58865 & 94267.30 & 7759 \\
        \bottomrule
        \end{tabular}
        \captionsetup{width=.9\linewidth}
        \caption{Number of posting accounts with mean number of followers.}
        \label{tab:followers}
\end{table}

We also analyze whether anti and pro accounts differ in their verification rates.
Verified accounts on Twitter are those which have been identified by the platform to be authentic and of public interest.
The criteria Twitter uses to grant verification badges to accounts are that the account must be ``authentic, notable, and active''.\footnote{\href{https://help.twitter.com/en/managing-your-account/about-twitter-verified-accounts}{Twitter: About verified accounts.}}
To compare verification rates, we filter out all retweets and retain only unique original authors.
We then perform a chi-square test of association which is significant, $\chi^2(1) = 945.07, \textit{p} < .001$.
Table \ref{tab:contingency} is the contingency table used in this test.
This suggests that pro-consensus accounts are significantly more likely to be verified than anti-consensus accounts.

\begin{table}[!t]
    \centering
    \small
    \begin{tabular}{lll|rr}
    \toprule
         & \multicolumn{2}{c}{\textbf{Anti}} & \multicolumn{2}{c}{\textbf{Pro}} \\
        \hline
        & \textit{Observed} & \textit{Expected} & \textit{Observed} & \textit{Expected} \\
        \midrule
        Unverified & 15055 & 14562.97 & 6861 & 7353.93 \\
        Verified & 312 & 804.03 & 898 & 405.97 \\
        \bottomrule
        \end{tabular}
        \caption{Contingency table for determining association between verification status and consensus stance.}
        \label{tab:contingency}
\end{table}

Next, we analyze popularity at the tweet level.
Our goal is to see whether pro-consensus tweets tend to get more retweets and likes, and whether this is commensurate with the followership of the accounts which post them.
For this purpose, we once again filter out retweets.
We do this because of the retweet object type in Twitter's API, which assigns to retweets the same number of retweets as the original tweet.
For example, if one tweet has 1000 retweets, then each of these retweets will also appear to have 1000 retweets of its own.
Therefore, we ignore retweets to avoid inflating popularity metrics.

At the tweet level, we examine five key metrics: number of likes, number of retweets, number of followers of posting account, follower-weighted likes ($\frac{likes + 1}{followers + 1}$) and follower-weighted retweets ($\frac{retweets + 1}{followers + 1}$).
The distribution of all metrics is heavily right-skewed mainly due to the presence of many zero values ($\textit{p} < .001$ in Shapiro-Wilks tests for all variables of interest).
For this reason, we utilize the Mann-Whitney U test which makes no parametric assumptions, for pairwise comparisons between anti and pro tweets.
The Mann-Whitney U test ranks values in an ordinal fashion from lowest to highest and performs comparisons based on this rank.
We report the results of these tests in Table \ref{tab:mann}. 

\begin{table}[t!]
    \centering
    \small
    \begin{tabular}{lcccccccr}
    \toprule
         & \multicolumn{2}{c}{\textbf{Anti}} & \multicolumn{2}{c}{\textbf{Pro}} & &  \\
        \hline
         & Mean rank & \textit{N} & Mean rank & \textit{N} & \textit{U} & \textit{Z} & \textit{p} \\
        \midrule
        Likes & 17050.88 & 25683 & 19321.71 & 9659 & 108096706.0 & 20.85 & < .001 \\
        Retweets (RTs) & 17142.18 & 25683 & 19078.93 & 9659 & 110441637.5 & 20.41 & < .001 \\
        Followers & 16202.31 & 25683 & 21578.02 & 9659 & 86302965.5 & 44.15 & < .001 \\
        Likes per follower & 19024.54 & 25683 & 14073.79 & 9659 & 89285802.5 & 40.66 & < .001 \\
        RTs per follower & 19133.76 & 25683 & 13783.40 & 9659 & 86480900.5 & 43.94 & < .001 \\
        \bottomrule
        \end{tabular}
        \caption{Results of Mann Whitney U tests. Bonferroni corrections are applied.}
        \label{tab:mann}
\end{table}

A higher mean rank means the values in that group are also higher overall.
As is seen in Table~\ref{tab:mann}, pro tweets receive significantly higher likes and retweets than anti tweets and are posted by accounts with significantly higher numbers of followers.
However, when we look at popularity metrics relative to the number of followers of the posting accounts, the opposite is true.
Anti posts receive more likes and retweets per follower than pro posts.
This suggests that, while pro posters are generally more popular on Twitter, anti tweets garner more relative attention and drive higher engagement.

\section{Social network analysis}\label{sec:sna}

Our third and final research question concerned the topology of the network which propagates pro-consensus and anti-consensus information across the Twitter platform.
We constructed a weighted retweet network graph and visualized it using the ForceAtlas2 algorithm which clusters nodes based on how well-connected they are~\cite{jacomy_forceatlas2_2014}.
We applied the Louvain algorithm which is a modularity optimization method for community detection~\cite{blondel_fast_2008}, and colorized distinct communities based on their modularity class.
We performed these tasks using the Gephi software~\cite{bastian_gephi_2009}.

\subsection{General COVID-19 network visualization}
\begin{figure}[!htb]
    \centering
    \includegraphics[width=.8\textwidth]{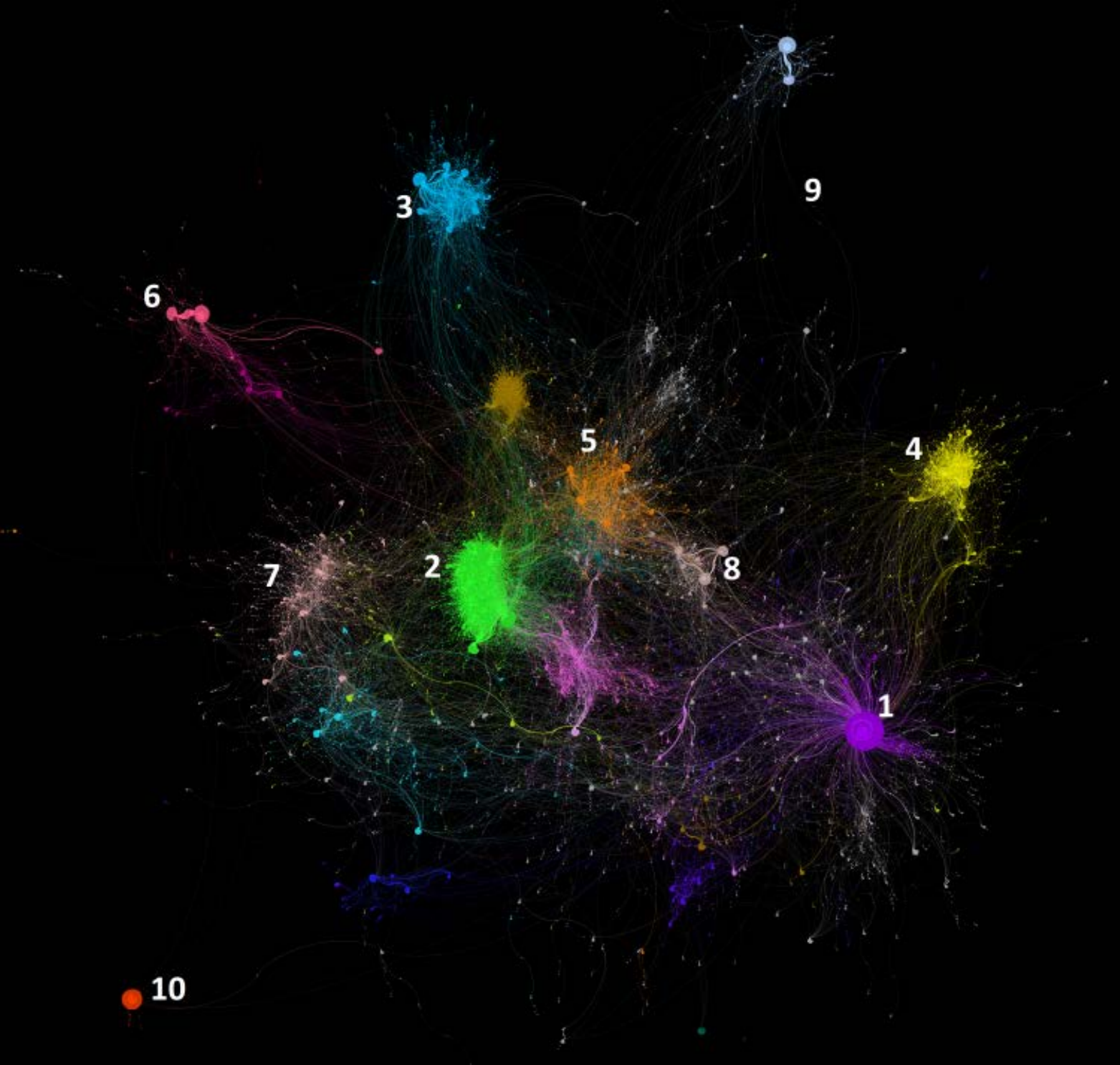}
    \caption{Retweet graph visualization for general COVID-19 dataset.}
    \label{fig:sna_mass}
\end{figure}

We first analyze the network graph for the general COVID-19 dataset.
We filter out any nodes which are not connected to the giant component (i.e. nodes which do not have any connections to the main network).
With this filtering, the network consists of 159K edges and 129K nodes.
This retweet network visualization is seen in Figure \ref{fig:sna_mass}, with the 10 largest communities labelled from biggest (1) to smallest (10).
We scale nodes based on in-degree.

\paragraph{First cluster.}The largest (purple) community contains 14.08\% of all network nodes.
This community is centered around the official Nobel prize account, which has by far the highest in-degree (18546) and highest authority (0.9999682) in the entire network.
The round shape of this community indicates that this central Nobel prize account is almost exclusively retweeted in this community (i.e., there is no information propagation between other nodes).
Its size also possibly stems from the traction that the official Nobel prize announcements receive.
Generally, this community almost exclusively discusses pro-consensus laureates.
Some other scientific accounts, such as the official Nature magazine account, also form part of this cluster.

\paragraph{Second cluster.}The second-largest (green community) contains 12.25\% of the nodes.
It is a French-based misinformation cluster which disproportionately mentions the anti-consensus laureate.
This community seems to bridge the entire network, since it contains the top two accounts in terms of betweenness centrality (126387.67 and 89829.35).
CNEWS, mentioned above as a key vehicle through which this laureate gained exposure, appears in this community.
It has the third-highest in-degree in the entire network (4206).
Accounts in this cluster are generally very prone to spreading misinformation.

\paragraph{Third cluster.}Next in terms of size is a Brazilian misinformation community (blue), which includes 6.89\% of network nodes.
Compared to the French misinformation cluster, the nodes here have fewer in-degrees but there are more ``central players''.
The entire community seems to be centered around an account which comes across as a news site, but shares unverified news.

\paragraph{Fourth cluster.}The fourth-biggest community (yellow, 5.69\% of nodes) is built around English-speaking mainstream media.
Examples of accounts appearing here are the BBC's Question Time (which often features scientists), and associated journalists. 
Fact-checking accounts such as FullFact also appear here, but are not central in the cluster.

\paragraph{Fifth cluster.}The orange cluster (5.32\% of network nodes) is an English-speaking misinformation community.
It contains several roughly equally-retweeted nodes with high in-degrees (highest being 1140).
Most accounts here systematically and frequently spread misinformation and COVID-19 conspiracy theories, with a few accounts seemingly dedicated to COVID-19 specifically.
A key node in this cluster is the account of a RAIR foundation journalist, possibly due to their role in popularizing the anti-consensus laureate's anti-vaccination sentiments.
This cluster also contains several accounts high in betweenness centrality, placing this cluster, along with the French misinformation cluster, in the middle of the entire network.
These two clusters also constitute the most important misinformation communities in the network.

\paragraph{Sixth cluster.}Another important community in the network is a Japanese misinformation cluster (reddish-pink, 4.72\% of nodes).
Nearly all of the network traffic concerns two accounts belonging to a book author and an academic (who has the fourth-highest in-degree in the network; 3572).
These accounts tweet both in Japanese and in English, and they mostly quote the anti-consensus laureate on his claims that the virus was engineered in a lab.
These accounts seem to be driven by a motivation to challenge China and call for accountability on its part.
They seem to use misinformation as a means of promoting this anti-China sentiment.

\paragraph{Seventh cluster.}Next, shown in beige-pink in the visualization, is an Argentine misinformation community which contains 4.32\% of the nodes found in the network.
This is a more loosely-connected community with no central players, however there do seem to be some accounts dedicated specifically to opposing the official story on COVID-19.

\paragraph{Eighth cluster.}We also observe another English-speaking misinformation community, which seems, however, to be based in India (light brown, 4.27\%).
Some accounts in this cluster exhibit anti-Muslim sentiment, however others do not seem to be systematic misinformers.
Rather, some accounts in this cluster shared the anti-consensus laureate's statements on a few occasions and drew a lot of retweets as a result.

\paragraph{Ninth cluster.}In light-blue is a Turkish misinformation community (4.11\% of network nodes).
The cluster seems to be centered around two journalist-like accounts, one of which has the fifth-highest in-degree in the network (3553).
These accounts behave like news aggregators and do not seem to misinform on purpose.
However, the majority of their traffic does stem from tweets concerning the anti-consensus laureate's statements, which were shared widely within the cluster.

\paragraph{Tenth cluster.}Finally, in red is an unrelated Japanese community containing 3.64\% of all network nodes.
There is only one central node in this cluster which also happens to be the second-highest in terms of in-degree (4715).
This account belongs to a Japanese boy-band whose members hosted a podcast episode with one of the pro-consensus laureates.
Nearly all of the traffic in this network comes from a single tweet which advertized this episode.
As seen from Figure \ref{fig:sna_mass}, this community is quite separated from the rest of the network, albeit still loosely connected.
This is likely due to the distinct audience that this account attracts, which is likely more interested in pop culture rather than the pandemic and surrounding policies.

\paragraph{Bot network locations.}We are also able to trace the top 5 relative posters that were identified as potential bots.
We use the top relative poster (User1) as a focal node and filter its ego network at a depth of 1 (i.e., accounts which have directly retweeted it).
We find that two of the remaining four accounts in the top 5 appear in this ego network (Users 0 and 2).  
When we extend the network depth to 3, we find that this botnet is situated between the French and English misinformation clusters (green and orange), with relatively close ties to the Brazilian misinformation cluster as well.
This suggests that information may propagate through this botnet across these three communities, although no nodes in this botnet show high centrality or authority metrics.
With regards to the second-highest relative user (User8), they are completely absent from the giant component of the network and thus inconsequential here.
The fifth-highest relative user (User3) is situated within the Argentine misinformation cluster.
It should be noted that the four nodes which are part of the giant component all belong to the same community (Argentine misinformation cluster), suggesting that they do retweet each other, but also amplify different types of accounts.

\subsection{COVID-19 vaccine network visualization.}

\begin{figure}[!htb]
    \centering
    \includegraphics[width=.8\textwidth]{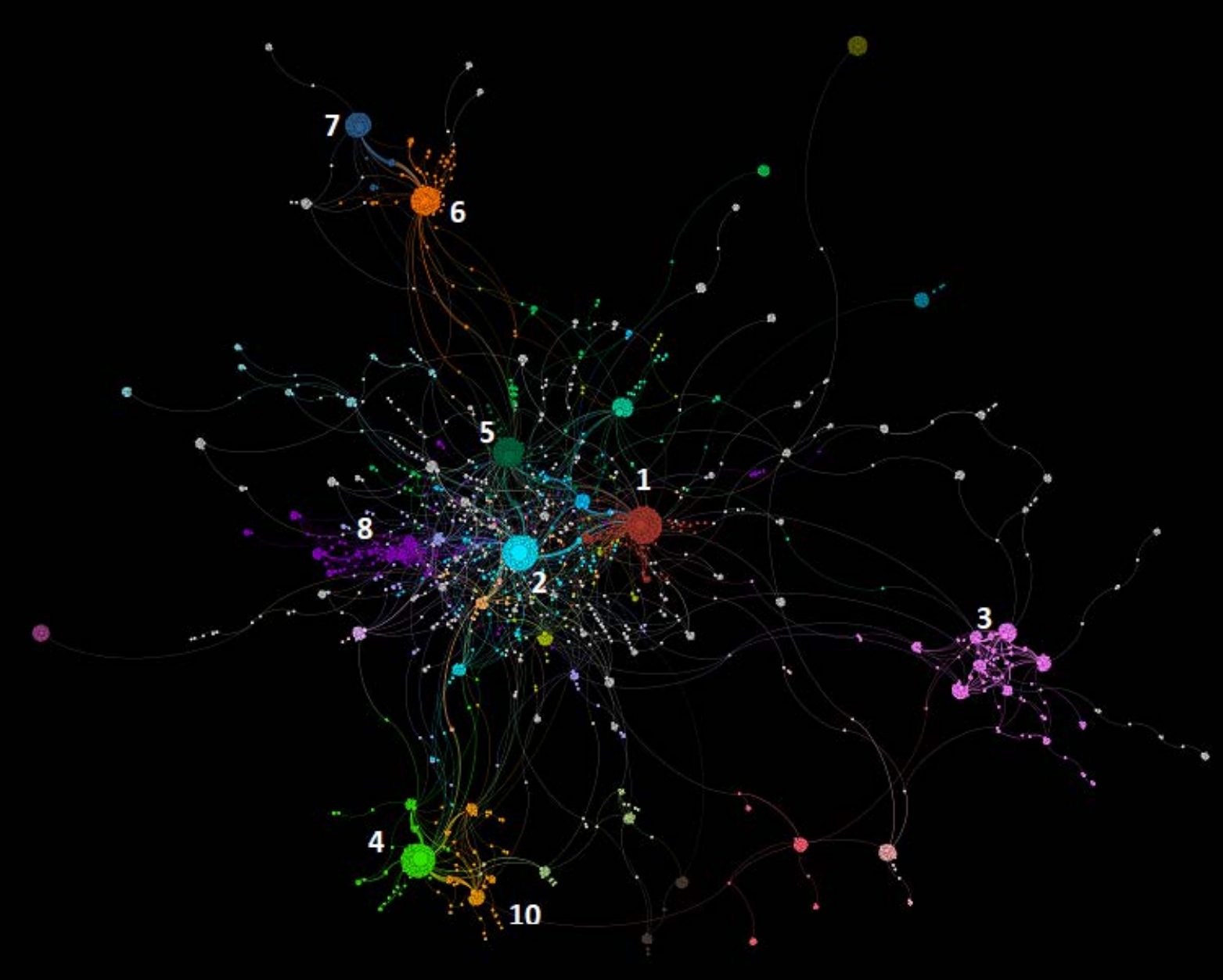}
    \caption{Retweet graph visualization for COVID-19 vaccine dataset.}
    \label{fig:sna_vac}
\end{figure}

We also visualize the retweet network graph for the vaccine dataset in Figure \ref{fig:sna_vac}, where the formation of communities is more straightforward.
We scale node size based on its in-degree.
The general pattern we observe is that most communities are centered around a single focal user, although some communities are more diverse in terms of their most retweeted users.
Again, we filter out any nodes which are not connected to the giant component, which leaves 9.8K nodes and 10.7K edges in the network.
As with the general COVID-19 network, we label communities from 1 (biggest) to 10 (10th biggest).
We note that the 9th biggest community is absent from the graph visualization because it is not connected to the giant component.

\paragraph{Anti-consensus communities with focal centers.}The two largest communities (red-brown (1), 8.51\% of nodes; light blue (2), 8.17\% of nodes) are centered around the RAIR foundation journalist discussed above and an account which tweets exclusively about COVID-19 conspiracy theories, respectively.
Both of these accounts appear in the English-speaking misinformation cluster of the general COVID-19 dataset (in orange, Figure \ref{fig:sna_mass}).
These two nodes are also the most important in the entire network.
The journalist has the highest betweenness centrality out of all nodes (4028.5) and the second-highest authority (0.302) and in-degree (970).
On the other hand, the conspiracy account has the highest authority (0.951) and in-degree (1026) in the network.

Communities 4-7 also follow similar trends.
Community 4 (6.66\%) is centered around a misinformation account with the third-highest in-degree in the network (833).
This account appears in the French misinformation cluster in the general COVID-19 network (in green, Figure \ref{fig:sna_mass}).
The fifth community (5.13\%) is centered around an account which similarly appears in the English-speaking misinformation community in the previous section.
This account has the fourth-highest in-degree in the network (650).
Communities 6 (4.95\%) and 7 (3.61\%) are also centered around two distinct nodes.
These concern Japanese accounts which are prone to misinformation, although they are not the same accounts which drive the traffic of the Japanese misinformation cluster in Figure \ref{fig:sna_mass}.

\paragraph{Decentralized anti-consensus communities.}Communities 8 (3.6\%) and 10 (2.6\%) also tend to amplify the anti-consensus laureate and generally spread misinformation, although they do not feature nodes which are as central as the previously discussed communities.
Community 8 tends to cross-tweet with community 2.
It features both Chinese and English-speaking accounts, making it a fairly suspicious cluster.
The most retweeted account here belongs to an independent Australian Member of Parliament who has shared many COVID-related conspiracy theories and amplified the anti-consensus laureate on several occasions. 
Community 10 is adjacent to community 4 as it is also a French-speaking, albeit less centralized, misinformation community which also amplifies the anti-consensus laureate.

\paragraph{Non-misinformation communities.}The third largest cluster (3) (containing 7.15\% of all nodes) belongs to a network of India-based fact-checkers who predominantly tweet in English.
Many of the most retweeted accounts in this community belong to members of an independent Indian organization dedicated to debunking online falsehoods.
Taking this into account, it is possible that the high betweenness centrality of the RAIR foundation journalist stems from the fact that they are connected both to all misinformation clusters, but also to this fact-checking group which targeted their anti-vaccine piece on the anti-consensus laureate.

The ninth-biggest community (3.06\% of nodes) also seems to be fairly pro-science, featuring the official Nobel prize account and medical doctors, among others (although we do note that some medical doctors are featured in misinformation clusters where they are seen amplifying the anti-consensus laureate).
However, this community is completely separate from the giant component of the network and therefore does not appear in Figure \ref{fig:sna_vac}.

\section{Discussion}

In this work, we examined whether consensus-dissenting scientists are disproportionately amplified on Twitter such that a false consensus effect is created around COVID-19.

\subsection{Summary of findings}

We used a proxy for scientific consensus consisting of Nobelists in Physiology or Medicine.
Using this method, we derived the scientific consensus ratio of anti-vaccine to pro-vaccine scientists as 1:24.
We expanded this ratio to broader COVID-19 topics as well, beyond just vaccines.

\subsubsection{False consensus trends}

Our results showed that between January 2020 and July 2021, the anti-consensus (minority) scientist was amplified by a factor of 426.05 relative to the scientific consensus ratio in the context of COVID-19 vaccines, and by a factor of 42.9 in the context of COVID-19 in general.
These findings cannot be explained by the previous popularity of consensus-abiding or dissenting scientists, since baseline data collected prior to COVID-19 becoming a trending topic suggested that Twitter conversations roughly followed the scientific consensus ratio.
In other words, we observed a strong false consensus effect on the Twitter platform in COVID-19-related topics and especially on COVID-19 vaccines.

Time-series analyses revealed that mentions of the laureates we used in our ratio are predominantly event-driven.
The consensus-abiding laureates received the most mentions around the time of their announcement as Nobel prize winners.
However, the dissenting laureate received the most attention on two occasions where they made consensus-opposing claims to mainstream and alternative media.
Whereas we were able to find several occasions where other laureates made media statements during our consensus categorization exercise (Appendix~\ref{append_a}), there were no commensurate spikes following these appearances.
This suggests that misinformation-laden yet scientific-sounding information is uniquely equipped to propagate through social media, consistent with the findings of~\cite{loomba_measuring_2021} on the believability of ``scientific'' misinformation.

\subsubsection{Popularity analyses}

We also examined the drivers of this false consensus on Twitter from a popularity and engagement perspective.
We found that the top 20 anti-consensus posters produce almost 4.5 times as much content as the top 20 pro-consensus posters.
However, we also found that on average, top posters in both categories exhibited quite high bot-like activity (as expected, since a lot of these may be aggregator bots).
In terms of potential for reach, we found that pro-consensus accounts were generally more likely to be verified, and that they amassed larger followerships than anti-accounts.
In addition, pro-consensus tweets did receive more likes and retweets than anti-consensus tweets.
However, when normalizing engagement by the posting accounts' potential for reach (number of followers), we found that anti-consensus posts received higher \textit{relative} engagement (i.e., more likes and retweets per follower).
This suggests that the dissenting scientist mainly receives exposure through higher engagement and more relative propagation, rather than due to being featured on more popular accounts than the consensus-abiding scientists.

\subsubsection{Retweet network analyses}

We conducted network analyses on both COVID-19 vaccines specifically and COVID-19 in general.
For COVID-19 more generally, we were able to find that among the 10 largest communities, 7 communities are misinformation prone.
These communities span multiple regions, namely Japan, Brazil, Argentina, Turkey, France, India, and English-speaking countries.

Contrary to Singh et al.~\cite{singh_understanding_2020}, we did not find that mainstream news sources are centrally located in the network.
Rather, the main brokerage role seems to fall with the French misinformation cluster which also includes the nodes with the highest betweenness centralities in the entire network.
The English-speaking misinformation cluster is also relatively quite central.
We believe the difference between our results and \cite{singh_understanding_2020} stems from the fact that Singh et al. relied on a URL domain approach to collect and categorize misinformation and reliable news.
Since we used names of featured scientists to do this, our results may suggest that mainstream media do not feature consensus-abiding scientists or do not receive enough exposure when they do so in order to effectively intercept such a network.
However, this would need to be verified with a wider set of scientists making up the scientific consensus.

With respect to the vaccine retweet network, the dynamics change slightly.
Once again, this network is dominated by misinformation-prone clusters which amplify the dissenting scientist (8 out of the 10 largest communities).
However, communities tend to be structured around a central node which is usually the community ``leader'' in terms of in-degrees.
This is the case for all but two of the false consensus-driving communities.

Strangely, we also observe that the ninth-largest community, featuring mainly pro-consensus accounts, is fully disconnected from the main component.
In the context of COVID-19 vaccines, this suggests that the misinformation communities do not receive any exposure to the true scientific consensus from this community whatsoever.
Nonetheless, we find that the third largest community in the vaccine network is dedicated to fact-checking mainly the largest misinformation community, placing the latter in the middle of the entire network in terms of betweenness centrality.

\subsection{Implications}

This work carries several insights for practice.
To our knowledge, it is the first paper to examine false consensus biases in the context of COVID-19.
It is also the first paper to study deviations between public perception and scientific consensus using social media data.
Previously, Harvey~\cite{harvey_internet_2018} was able to demonstrate that approximately 80\% of the 90 climate change-denying blogs in their sample cited the same professor as the main source for their claims.
Here we show a similar pattern regarding the topic of COVID-19, since the one scientist who dissented from scientific consensus in our ratio was also the one who was disproportionately amplified.
This is important because prior research suggests that the average person is fixated on the number of secondary, and not primary sources for a claim to assess its credibility~\cite{yousif_illusion_2019}.
In other words, even though the dissenting scientist (primary source in our case) was a very small minority, the fact that multiple secondary sources (other accounts) cited them is likely to have created the illusion that their view is much more established in the scientific community than it really is.
This may be especially true for misinformation-prone, homogeneous groups which already hold anti-vaccination or otherwise conspiratorial sentiments~\cite{wojcieszak_false_2008}.

Overall, we find that dissenting scientific voices are amplified to a much larger extent than consensus-abiding scientific voices.
We thus identify dissenting scientific voices as a prominent source of COVID-19 misinformation on Twitter.  
Regarding social media governance around misinformation, a worthwhile avenue of controlling the spread of false information particularly around COVID-19 is an attempt to highlight the voices of consensus-abiding scientists.
Especially because secondary sources can play a major role in shaping the informational ecosystem~\cite{yousif_illusion_2019}, balancing out the amplification of dissenting voices with commensurate exposure to scientists who subscribe to true consensus may serve to at least reduce the rate at which scientific-sounding misinformation propagates.

Consistent with findings around the relative propagation of true and false information on Twitter~\cite{vosoughi_spread_2018} we also find that consensus-dissenting tweets receive higher relative engagement compared to consensus-abiding tweets, even though the latter tend to be posted by more recognized and popular accounts.
It is likely that consensus-dissenting opinions, even when stemming from scientists, may serve other purposes such as legitimizing anti-establishment sentiments for the user amplifying them~\cite{enders_different_2020,pickles_covid-19_2021,roozenbeek_susceptibility_2020}.
Therefore, a worthwhile avenue of further combating misinformation may be addressing the dissenting voices head-on instead of simply promoting content from scientists subscribing to scientific consensus in a disassociated manner.
This could be in the form of consulting consensus-abiding voices to ``fact-check'' consensus-dissenting voices, such that the normalization of relative exposure (and thus the reduction of false consensus) becomes inherent to the mechanisms via which scientific disagreement takes place on social media.

With respect to our social network analyses, we find that events which grant exposure to consensus-dissenting scientists are propagated internationally.
We only observe spikes in mentions around the anti-consensus laureate periodically (Figure~\ref{fig:comparison}) following media appearances, however these events are discussed on Twitter across multiple countries.
This means that scientist-backed misinformation, even when produced locally in one country as was the case with the anti-consensus laureate in question, can have wide-reaching resonance.
To this end, early detection and control of disproportionate amplification can be impactful.
However, a temporal analysis of the retweet network development could provide additional insights as to exactly how useful such an approach would be.

\subsection{Limitations and future research directions}

The present work does suffer from some limitations.
First and foremost, our proxy method for determining scientific consensus was quite restricted.
Although there are thousands of scientists working on various aspects of the COVID-19 pandemic, we used only 25 people based on our selection criteria to form a ratio of scientific consensus.
It is very likely that the false consensus factors we report here would vastly change based on a different, more extensive ratio.
We therefore call on future research to either collate information on true scientific consensus around the COVID-19 pandemic (much how~\cite{cook_quantifying_2013} did for scientific consensus on climate change), or to utilize exploratory computational methods to derive this based on named-entity recognition.

Furthermore, we based our tweet classification approach on a fairly unsophisticated heuristic rule.
Future research could utilize natural language processing models such as BERT~\cite{devlin_bert_2019} to more accurately classify tweets into anti- or pro-consensus, due to instances of the heuristic rule failing as described in~\S\ref{sec:study}.
This would also assist with better context recognition to prevent the introduction of synonymy-induced noise (i.e., identifying whether a tweet is truly talking about a specific Nobel laureate and not another person with the same name).
However, future research could also utilize similar approaches in order to detect instances where broader COVID-19 misinformation on Twitter and other social media platforms invokes the opinion of scientifically credible figures.

Finally, this paper reports on an exploratory analysis of the false consensus information ecosystem on COVID-19 on Twitter.
For this reason, we do not go into depth regarding the exact linguistic attributes of tweets which cite consensus-abiding or consensus-dissenting scientists.
Such an analysis in future research would be promising, as it could reveal textual features which may be driving the increased engagement of anti-consensus tweets relative to the pro-consensus tweets.

\subsection{Conclusion}

In this study, we operationalized false consensus as the deviation between public consensus on Twitter and ``true'' scientific consensus as derived through our proxy method.
We are able to answer our research questions as follows.

\begin{enumerate}
    \item[\textbf{RQ1:}] Dissenting scientists are indeed amplified to a large extent relative to consensus-abiding scientists. Specifically, we observe a false consensus factor of 426.05 in the context of COVID-19 vaccines, and a factor of 42.9 in the context of COVID-19 more generally.
    \item[\textbf{RQ2:}] Consensus-dissenting tweets seem to drive higher engagement than consensus-abiding tweets. Specifically, we find that pro-consensus tweets tend to be produced by more popular and recognized accounts. However, relative to their potential for reach, anti-consensus tweets receive more likes and retweets per potential user (follower) who could be exposed to them.
    \item[\textbf{RQ3:}] For COVID-19 in general, consensus-dissenting tweets propagate across the platform to reach audiences in several different countries. Anti-consensus clusters are more centrally located than pro-consensus clusters, which tend to be more mainstream, yet more marginalized in this specific context. With respect to COVID-19 vaccines, clusters are predominantly centered around focal users whose amplifications of consensus-dissenting scientists receive a lot of engagement. We do find some attempts to revert this as a certain community targets these high-engagement tweets with debunking and fact-checking efforts.
\end{enumerate}

We hope that this research can inform social media governance and response to misinformation, as well as highlighting the use of credible messengers~\cite{dolan_influencing_2012} as a method of debunking scientific-sounding falsehoods and balancing out false consensus.

\section{Acknowledgments}
This research has been partially funded by the UK EPSRC grant EP/S022503/1 which supports the UCL Centre for Doctoral Training in Cybersecurity. Any opinions, conclusions, or recommendations expressed in this work are those of the authors and do not necessarily reflect the views of the UK EPSRC.

\newpage
\bibliographystyle{acm}
\bibliography{refs}  

\begin{thebibliography}{10}

\bibitem{noauthor_general_nodate}
General {Data} {Protection} {Regulation} ({GDPR}) {Compliance} {Guidelines}.

\bibitem{noauthor_who_2021}
{WHO} {Coronavirus} ({COVID}-19) {Dashboard}, 2021.

\bibitem{al-rakhami_lies_2020}
{\sc Al-Rakhami, M.~S., and Al-Amri, A.~M.}
\newblock Lies {Kill}, {Facts} {Save}: {Detecting} {COVID}-19 {Misinformation}
  in {Twitter}.
\newblock {\em IEEE Access 8\/} (2020), 155961--155970.
\newblock Conference Name: IEEE Access.

\bibitem{allen_climate_2014}
{\sc Allen, S.~K., Plattner, G.-K., Nauels, A., Xia, Y., and Stocker, T.~F.}
\newblock Climate {Change} 2013: {The} {Physical} {Science} {Basis}. {An}
  overview of the {Working} {Group} 1 contribution to the {Fifth} {Assessment}
  {Report} of the {Intergovernmental} {Panel} on {Climate} {Change} ({IPCC}).
\newblock 3544.
\newblock Conference Name: EGU General Assembly Conference Abstracts.

\bibitem{alwan_scientific_2020}
{\sc Alwan, N.~A., Burgess, R.~A., Ashworth, S., Beale, R., Bhadelia, N.,
  Bogaert, D., Dowd, J., Eckerle, I., Goldman, L.~R., Greenhalgh, T.,
  Gurdasani, D., Hamdy, A., Hanage, W.~P., Hodcroft, E.~B., Hyde, Z., Kellam,
  P., Kelly-Irving, M., Krammer, F., Lipsitch, M., McNally, A., McKee, M.,
  Nouri, A., Pimenta, D., Priesemann, V., Rutter, H., Silver, J., Sridhar, D.,
  Swanton, C., Walensky, R.~P., Yamey, G., and Ziauddeen, H.}
\newblock Scientific consensus on the {COVID}-19 pandemic: we need to act now.
\newblock {\em The Lancet 396}, 10260 (Oct. 2020), e71--e72.
\newblock Publisher: Elsevier.

\bibitem{badell-grau_investigating_2020}
{\sc Badell-Grau, R.~A., Cuff, J.~P., Kelly, B.~P., Waller-Evans, H., and
  Lloyd-Evans, E.}
\newblock Investigating the {Prevalence} of {Reactive} {Online} {Searching} in
  the {COVID}-19 {Pandemic}: {Infoveillance} {Study}.
\newblock {\em Journal of Medical Internet Research 22}, 10 (Oct. 2020),
  e19791.
\newblock Company: Journal of Medical Internet Research Distributor: Journal of
  Medical Internet Research Institution: Journal of Medical Internet Research
  Label: Journal of Medical Internet Research Publisher: JMIR Publications
  Inc., Toronto, Canada.

\bibitem{bastian_gephi_2009}
{\sc Bastian, M., Heymann, S., and Jacomy, M.}
\newblock Gephi: {An} {Open} {Source} {Software} for {Exploring} and
  {Manipulating} {Networks}.
\newblock In {\em Third {International} {AAAI} {Conference} on {Weblogs} and
  {Social} {Media}\/} (Mar. 2009).

\bibitem{bavel_using_2020}
{\sc Bavel, J. J.~V., Baicker, K., Boggio, P.~S., Capraro, V., Cichocka, A.,
  Cikara, M., Crockett, M.~J., Crum, A.~J., Douglas, K.~M., Druckman, J.~N.,
  Drury, J., Dube, O., Ellemers, N., Finkel, E.~J., Fowler, J.~H., Gelfand, M.,
  Han, S., Haslam, S.~A., Jetten, J., Kitayama, S., Mobbs, D., Napper, L.~E.,
  Packer, D.~J., Pennycook, G., Peters, E., Petty, R.~E., Rand, D.~G., Reicher,
  S.~D., Schnall, S., Shariff, A., Skitka, L.~J., Smith, S.~S., Sunstein,
  C.~R., Tabri, N., Tucker, J.~A., Linden, S. v.~d., Lange, P.~v., Weeden,
  K.~A., Wohl, M. J.~A., Zaki, J., Zion, S.~R., and Willer, R.}
\newblock Using social and behavioural science to support {COVID}-19 pandemic
  response.
\newblock {\em Nature Human Behaviour 4}, 5 (May 2020), 460--471.
\newblock Bandiera\_abtest: a Cg\_type: Nature Research Journals Number: 5
  Primary\_atype: Reviews Publisher: Nature Publishing Group Subject\_term:
  Human behaviour;Immunology;Sociology Subject\_term\_id:
  human-behaviour;immunology;sociology.

\bibitem{bayes_research_2020}
{\sc Bayes, R., Bolsen, T., and Druckman, J.~N.}
\newblock A {Research} {Agenda} for {Climate} {Change} {Communication} and
  {Public} {Opinion}: {The} {Role} of {Scientific} {Consensus} {Messaging} and
  {Beyond}.
\newblock {\em Environmental Communication\/} (Sept. 2020), 1--19.

\bibitem{blondel_fast_2008}
{\sc Blondel, V.~D., Guillaume, J.-L., Lambiotte, R., and Lefebvre, E.}
\newblock Fast unfolding of communities in large networks.
\newblock {\em Journal of Statistical Mechanics: Theory and Experiment 2008},
  10 (Oct. 2008), P10008.

\bibitem{bridgman_causes_2020}
{\sc Bridgman, A., Merkley, E., Loewen, P.~J., Owen, T., Ruths, D., Teichmann,
  L., and Zhilin, O.}
\newblock The causes and consequences of {COVID}-19 misperceptions:
  {Understanding} the role of news and social media.
\newblock {\em Harvard Kennedy School Misinformation Review 1}, 3 (June 2020).

\bibitem{cha_prevalence_2021}
{\sc Cha, M., Cha, C., Singh, K., Lima, G., Ahn, Y.-Y., Kulshrestha, J., and
  Varol, O.}
\newblock Prevalence of {Misinformation} and {Factchecks} on the {COVID}-19
  {Pandemic} in 35 {Countries}: {Observational} {Infodemiology} {Study}.
\newblock {\em JMIR Human Factors 8}, 1 (Feb. 2021), e23279.
\newblock Company: JMIR Human Factors Distributor: JMIR Human Factors
  Institution: JMIR Human Factors Label: JMIR Human Factors Publisher: JMIR
  Publications Inc., Toronto, Canada.

\bibitem{cook_quantifying_2013}
{\sc Cook, J., Nuccitelli, D., Green, S.~A., Richardson, M., Winkler, B.,
  Painting, R., Way, R., Jacobs, P., and Skuce, A.}
\newblock Quantifying the consensus on anthropogenic global warming in the
  scientific literature.
\newblock {\em Environmental Research Letters 8}, 2 (May 2013), 024024.
\newblock Publisher: IOP Publishing.

\bibitem{davis_botornot_2016}
{\sc Davis, C.~A., Varol, O., Ferrara, E., Flammini, A., and Menczer, F.}
\newblock {BotOrNot}: {A} {System} to {Evaluate} {Social} {Bots}.
\newblock In {\em Proceedings of the 25th {International} {Conference}
  {Companion} on {World} {Wide} {Web}\/} (Republic and Canton of Geneva, CHE,
  Apr. 2016), {WWW} '16 {Companion}, International World Wide Web Conferences
  Steering Committee, pp.~273--274.

\bibitem{devlin_bert_2019}
{\sc Devlin, J., Chang, M.-W., Lee, K., and Toutanova, K.}
\newblock {BERT}: {Pre}-training of {Deep} {Bidirectional} {Transformers} for
  {Language} {Understanding}.
\newblock In {\em Proceedings of the 2019 {Conference} of the {North}
  {American} {Chapter} of the {Association} for {Computational} {Linguistics}:
  {Human} {Language} {Technologies}, {Volume} 1 ({Long} and {Short}
  {Papers})\/} (Minneapolis, Minnesota, June 2019), Association for
  Computational Linguistics, pp.~4171--4186.

\bibitem{dolan_influencing_2012}
{\sc Dolan, P., Hallsworth, M., Halpern, D., King, D., Metcalfe, R., and Vlaev,
  I.}
\newblock Influencing behaviour: {The} mindspace way.
\newblock {\em Journal of Economic Psychology 33}, 1 (Feb. 2012), 264--277.

\bibitem{enders_different_2020}
{\sc Enders, A.~M., Uscinski, J.~E., Klofstad, C., and Stoler, J.}
\newblock The different forms of {COVID}-19 misinformation and their
  consequences.
\newblock {\em The Harvard Kennedy School Misinformation Review\/} (Nov. 2020).
\newblock Accepted: 2020-12-03T10:01:37Z Publisher: Shorenstein Center for
  Media, Politics and Public Policy, at Harvard University, John F. Kennedy
  School of Government.

\bibitem{gallacher_mutual_2021}
{\sc Gallacher, J.~D., and Heerdink, M.}
\newblock Mutual radicalisation of opposing extremist groups via the
  {Internet}.
\newblock preprint, PsyArXiv, Feb. 2021.

\bibitem{goldberg_social_2020}
{\sc Goldberg, M., Gustafson, A., Maibach, E., Linden, S. v.~d., Ballew, M.~T.,
  Bergquist, P., Kotcher, J., Marlon, J.~R., Rosenthal, S.~A., and Leiserowitz,
  A.}
\newblock Social norms motivate {COVID}-19 preventive behaviors.
\newblock preprint, PsyArXiv, May 2020.

\bibitem{harvey_internet_2018}
{\sc Harvey, J.~A., van~den Berg, D., Ellers, J., Kampen, R., Crowther, T.~W.,
  Roessingh, P., Verheggen, B., Nuijten, R. J.~M., Post, E., Lewandowsky, S.,
  Stirling, I., Balgopal, M., Amstrup, S.~C., and Mann, M.~E.}
\newblock Internet {Blogs}, {Polar} {Bears}, and {Climate}-{Change} {Denial} by
  {Proxy}.
\newblock {\em BioScience 68}, 4 (Apr. 2018), 281--287.

\bibitem{havey_partisan_2020}
{\sc Havey, N.~F.}
\newblock Partisan public health: how does political ideology influence support
  for {COVID}-19 related misinformation?
\newblock {\em Journal of Computational Social Science 3}, 2 (Nov. 2020),
  319--342.

\bibitem{jacomy_forceatlas2_2014}
{\sc Jacomy, M., Venturini, T., Heymann, S., and Bastian, M.}
\newblock {ForceAtlas2}, a {Continuous} {Graph} {Layout} {Algorithm} for
  {Handy} {Network} {Visualization} {Designed} for the {Gephi} {Software}.
\newblock {\em PLOS ONE 9}, 6 (June 2014), e98679.
\newblock Publisher: Public Library of Science.

\bibitem{kouzy_coronavirus_2020}
{\sc Kouzy, R., Abi~Jaoude, J., Kraitem, A., El~Alam, M.~B., Karam, B., Adib,
  E., Zarka, J., Traboulsi, C., Akl, E.~W., and Baddour, K.}
\newblock Coronavirus {Goes} {Viral}: {Quantifying} the {COVID}-19
  {Misinformation} {Epidemic} on {Twitter}.
\newblock {\em Cureus 12}, 3 (2020), e7255.

\bibitem{lee_associations_2020}
{\sc Lee, J.~J., Kang, K.-A., Wang, M.~P., Zhao, S.~Z., Wong, J. Y.~H.,
  O'Connor, S., Yang, S.~C., and Shin, S.}
\newblock Associations {Between} {COVID}-19 {Misinformation} {Exposure} and
  {Belief} {With} {COVID}-19 {Knowledge} and {Preventive} {Behaviors}:
  {Cross}-{Sectional} {Online} {Study}.
\newblock {\em Journal of Medical Internet Research 22}, 11 (Nov. 2020),
  e22205.
\newblock Company: Journal of Medical Internet Research Distributor: Journal of
  Medical Internet Research Institution: Journal of Medical Internet Research
  Label: Journal of Medical Internet Research Publisher: JMIR Publications
  Inc., Toronto, Canada.

\bibitem{lees_intentions_2020}
{\sc Lees, J., Cetron, J.~S., Vollberg, M.~C., Reggev, N., and Cikara, M.}
\newblock Intentions to comply with {COVID}-19 preventive behaviors are
  associated with personal beliefs, independent of perceived social norms.
\newblock Tech. rep., PsyArXiv, May 2020.
\newblock type: article.

\bibitem{leviston_your_2013}
{\sc Leviston, Z., Walker, I., and Morwinski, S.}
\newblock Your opinion on climate change might not be as common as you think.
\newblock {\em Nature Climate Change 3}, 4 (Apr. 2013), 334--337.

\bibitem{lewandowsky_pivotal_2013}
{\sc Lewandowsky, S., Gignac, G.~E., and Vaughan, S.}
\newblock The pivotal role of perceived scientific consensus in acceptance of
  science.
\newblock {\em Nature Climate Change 3}, 4 (Apr. 2013), 399--404.

\bibitem{linden_scientific_2015}
{\sc Linden, S. L. v.~d., Leiserowitz, A.~A., Feinberg, G.~D., and Maibach,
  E.~W.}
\newblock The {Scientific} {Consensus} on {Climate} {Change} as a {Gateway}
  {Belief}: {Experimental} {Evidence}.
\newblock {\em PLOS ONE 10}, 2 (Feb. 2015), e0118489.
\newblock Publisher: Public Library of Science.

\bibitem{lockyer_understanding_2021}
{\sc Lockyer, B., Islam, S., Rahman, A., Dickerson, J., Pickett, K., Sheldon,
  T., Wright, J., McEachan, R., and Sheard, L.}
\newblock Understanding {COVID}-19 misinformation and vaccine hesitancy in
  context: {Findings} from a qualitative study involving citizens in
  {Bradford}, {UK}.
\newblock {\em Health Expectations n/a}, n/a (2021).
\newblock \_eprint: https://onlinelibrary.wiley.com/doi/pdf/10.1111/hex.13240.

\bibitem{loomba_measuring_2021}
{\sc Loomba, S., de~Figueiredo, A., Piatek, S.~J., de~Graaf, K., and Larson,
  H.~J.}
\newblock Measuring the impact of {COVID}-19 vaccine misinformation on
  vaccination intent in the {UK} and {USA}.
\newblock {\em Nature Human Behaviour 5}, 3 (Mar. 2021), 337--348.
\newblock Bandiera\_abtest: a Cg\_type: Nature Research Journals Number: 3
  Primary\_atype: Research Publisher: Nature Publishing Group Subject\_term:
  Science, technology and society;Social policy Subject\_term\_id:
  science-technology-and-society;social-policy.

\bibitem{lucia_covid-19_2020}
{\sc Lucia, V.~C., Kelekar, A., and Afonso, N.~M.}
\newblock {COVID}-19 vaccine hesitancy among medical students.
\newblock {\em Journal of Public Health (Oxford, England)\/} (Dec. 2020),
  fdaa230.

\bibitem{malik_determinants_2020}
{\sc Malik, A.~A., McFadden, S.~M., Elharake, J., and Omer, S.~B.}
\newblock Determinants of {COVID}-19 vaccine acceptance in the {US}.
\newblock {\em EClinicalMedicine 26\/} (Sept. 2020), 100495.

\bibitem{memon_characterizing_2020}
{\sc Memon, S.~A., and Carley, K.~M.}
\newblock Characterizing {COVID}-19 {Misinformation} {Communities} {Using} a
  {Novel} {Twitter} {Dataset}.
\newblock {\em arXiv:2008.00791 [cs]\/} (Sept. 2020).
\newblock arXiv: 2008.00791.

\bibitem{murphy_psychological_2021}
{\sc Murphy, J., Vallières, F., Bentall, R.~P., Shevlin, M., McBride, O.,
  Hartman, T.~K., McKay, R., Bennett, K., Mason, L., Gibson-Miller, J., Levita,
  L., Martinez, A.~P., Stocks, T. V.~A., Karatzias, T., and Hyland, P.}
\newblock Psychological characteristics associated with {COVID}-19 vaccine
  hesitancy and resistance in {Ireland} and the {United} {Kingdom}.
\newblock {\em Nature Communications 12}, 1 (Jan. 2021), 29.
\newblock Bandiera\_abtest: a Cc\_license\_type: cc\_by Cg\_type: Nature
  Research Journals Number: 1 Primary\_atype: Research Publisher: Nature
  Publishing Group Subject\_term: Human behaviour;Social sciences
  Subject\_term\_id: human-behaviour;social-sciences.

\bibitem{pickles_covid-19_2021}
{\sc Pickles, K., Cvejic, E., Nickel, B., Copp, T., Bonner, C., Leask, J.,
  Ayre, J., Batcup, C., Cornell, S., Dakin, T., Dodd, R.~H., Isautier, J.
  M.~J., and McCaffery, K.~J.}
\newblock {COVID}-19 {Misinformation} {Trends} in {Australia}: {Prospective}
  {Longitudinal} {National} {Survey}.
\newblock {\em Journal of Medical Internet Research 23}, 1 (Jan. 2021), e23805.
\newblock Company: Journal of Medical Internet Research Distributor: Journal of
  Medical Internet Research Institution: Journal of Medical Internet Research
  Label: Journal of Medical Internet Research Publisher: JMIR Publications
  Inc., Toronto, Canada.

\bibitem{robertson_predictors_2021}
{\sc Robertson, E., Reeve, K.~S., Niedzwiedz, C.~L., Moore, J., Blake, M.,
  Green, M., Katikireddi, S.~V., and Benzeval, M.~J.}
\newblock Predictors of {COVID}-19 vaccine hesitancy in the {UK} household
  longitudinal study.
\newblock {\em Brain, Behavior, and Immunity 94\/} (May 2021), 41--50.

\bibitem{roozenbeek_susceptibility_2020}
{\sc Roozenbeek, J., Schneider, C.~R., Dryhurst, S., Kerr, J., Freeman, A.
  L.~J., Recchia, G., van~der Bles, A.~M., and van~der Linden, S.}
\newblock Susceptibility to misinformation about {COVID}-19 around the world.
\newblock {\em Royal Society Open Science 7}, 10 (2020), 201199.
\newblock Publisher: Royal Society.

\bibitem{ross_false_1977}
{\sc Ross, L., Greene, D., and House, P.}
\newblock The “false consensus effect”: {An} egocentric bias in social
  perception and attribution processes.
\newblock {\em Journal of Experimental Social Psychology 13}, 3 (May 1977),
  279--301.

\bibitem{salali_covid-19_2020}
{\sc Salali, G.~D., and Uysal, M.~S.}
\newblock {COVID}-19 vaccine hesitancy is associated with beliefs on the origin
  of the novel coronavirus in the {UK} and {Turkey}.
\newblock {\em Psychological Medicine\/} (Oct. 2020), 1--3.

\bibitem{sallam_covid-19_2021}
{\sc Sallam, M.}
\newblock {COVID}-19 {Vaccine} {Hesitancy} {Worldwide}: {A} {Concise}
  {Systematic} {Review} of {Vaccine} {Acceptance} {Rates}.
\newblock {\em Vaccines 9}, 2 (Feb. 2021), 160.

\bibitem{schulz_we_2020}
{\sc Schulz, A., Wirth, W., and Müller, P.}
\newblock We {Are} the {People} and {You} {Are} {Fake} {News}: {A} {Social}
  {Identity} {Approach} to {Populist} {Citizens}’ {False} {Consensus} and
  {Hostile} {Media} {Perceptions}.
\newblock {\em Communication Research 47}, 2 (Mar. 2020), 201--226.
\newblock Publisher: SAGE Publications Inc.

\bibitem{shahi_exploratory_2021}
{\sc Shahi, G.~K., Dirkson, A., and Majchrzak, T.~A.}
\newblock An exploratory study of {COVID}-19 misinformation on {Twitter}.
\newblock {\em Online Social Networks and Media 22\/} (Mar. 2021), 100104.

\bibitem{shahsavari_conspiracy_2020}
{\sc Shahsavari, S., Holur, P., Wang, T., Tangherlini, T.~R., and Roychowdhury,
  V.}
\newblock Conspiracy in the time of corona: automatic detection of emerging
  {COVID}-19 conspiracy theories in social media and the news.
\newblock {\em Journal of Computational Social Science 3}, 2 (Nov. 2020),
  279--317.

\bibitem{singh_understanding_2020}
{\sc Singh, L., Bode, L., Budak, C., Kawintiranon, K., Padden, C., and Vraga,
  E.}
\newblock Understanding high- and low-quality {URL} {Sharing} on {COVID}-19
  {Twitter} streams.
\newblock {\em Journal of Computational Social Science 3}, 2 (Nov. 2020),
  343--366.

\bibitem{terren_echo_2021}
{\sc Terren, L., and Borge-Bravo, R.}
\newblock Echo {Chambers} on {Social} {Media}: {A} {Systematic} {Review} of the
  {Literature}.
\newblock {\em Review of Communication Research 9\/} (Mar. 2021), 99--118.

\bibitem{vosoughi_spread_2018}
{\sc Vosoughi, S., Roy, D., and Aral, S.}
\newblock The spread of true and false news online.
\newblock {\em Science 359}, 6380 (Mar. 2018), 1146--1151.

\bibitem{weinschenk_democratic_2021}
{\sc Weinschenk, A.~C., Panagopoulos, C., and Linden, S. v.~d.}
\newblock Democratic {Norms}, {Social} {Projection}, and {False} {Consensus} in
  the 2020 {U}.{S}. {Presidential} {Election}.
\newblock {\em Journal of Political Marketing 0}, 0 (July 2021), 1--14.
\newblock Publisher: Routledge \_eprint:
  https://doi.org/10.1080/15377857.2021.1939568.

\bibitem{wojcieszak_false_2008}
{\sc Wojcieszak, M.}
\newblock False {Consensus} {Goes} {Online}: {Impact} of {Ideologically}
  {Homogeneous} {Groups} on {False} {Consensus}.
\newblock {\em Public Opinion Quarterly 72}, 4 (Jan. 2008), 781--791.

\bibitem{yang_prevalence_2020}
{\sc Yang, K.-C., Torres-Lugo, C., and Menczer, F.}
\newblock Prevalence of {Low}-{Credibility} {Information} on {Twitter} {During}
  the {COVID}-19 {Outbreak}.
\newblock {\em arXiv:2004.14484 [cs]\/} (June 2020).
\newblock arXiv: 2004.14484.

\bibitem{yeager_moderation_2019}
{\sc Yeager, D.~S., Krosnick, J.~A., Visser, P.~S., Holbrook, A.~L., and Tahk,
  A.~M.}
\newblock Moderation of classic social psychological effects by demographics in
  the {U}.{S}. adult population: {New} opportunities for theoretical
  advancement.
\newblock {\em Journal of Personality and Social Psychology 117}, 6 (2019),
  e84.
\newblock Publisher: US: American Psychological Association.

\bibitem{yousif_illusion_2019}
{\sc Yousif, S.~R., Aboody, R., and Keil, F.~C.}
\newblock The {Illusion} of {Consensus}: {A} {Failure} to {Distinguish}
  {Between} {True} and {False} {Consensus}.
\newblock {\em Psychological Science 30}, 8 (Aug. 2019), 1195--1204.

\end{thebibliography}

\appendix

\section{Laureates included in ratio}\label{append_a}

Note that some of these laureates have expressed their opinions on more than one occasion, however we only give one source per laureate here.
We did not find any laureates who expressed conflicting opinions on separate occasions.

\begin{table}[h!]
    \centering
    \scalebox{0.75}{
    \begin{tabular}{ p{3.5cm} p{2cm} p{2cm} p{3cm} p{10cm} }
    \textbf{Laureate} & \textbf{Year awarded} & \textbf{Vaccine stance} & \textbf{Source} & \textbf{Source description} \\
    \hline
    Eric Kandel & 2000 & Pro & \href{https://peoplesvaccinealliance.medium.com/?p=e0589edd5704}{Open letter} & Signs open letter to expand vaccine access \\
    \hline
    Leland Hartwell & 2001 & Pro & \href{https://peoplesvaccinealliance.medium.com/?p=e0589edd5704}{Open letter} & Signs open letter to expand vaccine access \\
    \hline
    Paul Nurse & 2001 & Pro & \href{https://www.crick.ac.uk/news/2020-11-13_why-i-will-be-getting-a-covid-vaccine}{Statement} & Explains importance of vaccination \\
    \hline
    Tim hunt & 2001 & Pro & \href{https://peoplesvaccinealliance.medium.com/?p=e0589edd5704}{Open letter} & Signs open letter to expand vaccine access \\
    \hline
    Robert Horvitz & 2002 & Pro & \href{https://leaps.org/award-winning-scientists-offer-advice-to-the-next-president-of-the-united-states/h-robert-horvitz-ph-d}{Presidential advice} & Commends the efforts and speed in developing vaccines \\
    \hline
    Barry Marshall & 2005 & Pro & \href{https://www.news.uwa.edu.au/archive/2020060312130/research/uwa-expert-series-barry-marshall-confident-covid-19-vaccine-will-be-developed/}{Projection} & Predicts the vaccine manufacturing process \\
    \hline
    Andrew Fire & 2006 & Pro & \href{https://www.vice.com/en/article/7k9gya/stanford-scientists-reverse-engineer-moderna-vaccine-post-code-on-github}{Interview} & Promoted open science on vaccine technology \\
    \hline
    Robert Mello & 2006 & Pro & \href{https://boston.cbslocal.com/2021/02/19/covid-vaccine-messenger-rna-nobel-prize-winner-umass-memorial-researcher-craig-mello/}{Interview} & Describes mRNA vaccine technology as safe \\
    \hline
    Mario Capecchi & 2007 & Pro & \href{https://peoplesvaccinealliance.medium.com/?p=e0589edd5704}{Open letter} & Signs open letter to expand vaccine access \\
    \hline
    Francoise Barre-Sinoussi & 2008 & Pro & \href{https://peoplesvaccinealliance.medium.com/?p=e0589edd5704}{Open letter} & Signs open letter to expand vaccine access \\
    \hline
    Elizabeth Blackburn & 2009 & Pro & \href{https://peoplesvaccinealliance.medium.com/?p=e0589edd5704}{Open letter} & Signs open letter to expand vaccine access \\
    \hline
    Jules Hoffmann & 2011 & Pro & \href{http://www.usias.fr/en/news-events/news/single-news/article/un-geste-nobel-pour-la-vaccination-contre-le-covid-19/}{Public appearance} & Appears taking the vaccine and describes it as safe \\
    \hline
    Shinya Yamanaka & 2012 & Pro & \href{https://www.reuters.com/world/asia-pacific/japan-business-leaders-suggest-ways-govt-speed-up-vaccination-rate-2021-04-29/}{Interview} & Calls on the Japanese government to speed up vaccinations \\
    \hline
    Randy Schekman & 2013 & Pro & \href{https://news.cgtn.com/news/2020-12-14/Science-is-what-advances-civilizations-Wd4HMxqWqc/index.html}{Webinar} & Urges the public to follow official advice on vaccinations \\
    \hline
    John O'Keefe & 2014 & Pro & \href{https://peoplesvaccinealliance.medium.com/?p=e0589edd5704}{Open letter} & Signs open letter to expand vaccine access \\
    \hline
    May-Britt Moser & 2014 & Pro & \href{https://peoplesvaccinealliance.medium.com/?p=e0589edd5704}{Open letter} & Signs open letter to expand vaccine access \\
    \hline
    Edvard Moser & 2014 & Pro & \href{https://sciencebusiness.net/covid-19/news/viewpoint-covid-19-vaccines-sit-atop-mountain-early-stage-science}{Interview} & Commends early science which enabled quick vaccine development \\
    \hline
    Jeffrey Hall & 2017 & Pro & \href{https://peoplesvaccinealliance.medium.com/?p=e0589edd5704}{Open letter} & Signs open letter to expand vaccine access \\
    \hline
    Michael Rosbash & 2017 & Pro & \href{https://www.brandeis.edu/now/2020/may/rosbash-coronavirus-article.html}{Interview} & States that the quick development of a vaccine is crucial against COVID-19 \\
    \hline
    James Allison & 2018 & Pro & \href{https://twitter.com/MDAndersonNews/status/1346562833136443398}{Public appearance} & Appears taking the vaccine and declares ``proud to stand with science'' \\
    \hline
    Tasuku Honjo & 2018 & Pro & \href{https://peoplesvaccinealliance.medium.com/?p=e0589edd5704}{Open letter} & Signs open letter to expand vaccine access \\
    \hline
    Gregg Semenza & 2019 & Pro & \href{https://peoplesvaccinealliance.medium.com/?p=e0589edd5704}{Open letter} & Signs open letter to expand vaccine access \\
    \hline
    Michael Houghton & 2020 & Pro & \href{https://www.ualberta.ca/newtrail/research/michael-houghton-in-conversation.html}{Interview} & Urges people to take a vaccine as soon as it is available \\
    \hline
    Charles Rice & 2020 & Pro & \href{https://www.youtube.com/watch?v=93OpURJDJ34}{Projection} & Forecasts the quick development of vaccines \\
    \hline
    Luc Montagnier & 2008 & Anti & \href{https://rairfoundation.com/alert-luc-montagnier-did-not-say-vaccine-would-kill-people-in-two-years-heres-what-he-did-say-video/}{Interview} & Calls mass vaccination an ``unacceptable mistake'' \\
    \hline
    \end{tabular}}
    \label{tab:laureates}
\end{table}

\newpage

\section{Search query}\label{append_b}

\begin{table}[!h]
    \centering
    \small
    \begin{tabular}{c|c|c c}
    \textbf{Covid synonyms} & \textbf{Vaccine synonyms} & \textbf{Laureate names} & \\
    \hline
    covid & vaxx & Francoise Barre-Sinoussi & Randy Schekman \\
    covid-19 & vaccination & Eric Kandel & John O'Keefe \\
    coronavirus & vaccine & Leland Hartwell & May-Britt Moser \\
    rona & vac & Paul Nurse & Edvard Moser   \\
    pandemic & vaccinated & Tim Hunt  & Michael Rosbash \\
    plandemic & jab & Robert Horvitz & James Allison \\
    virus & vax & Barry Marshall  & Tasuku Honjo \\
    lockdown & anti-vax & Andrew Fire & Gregg Semenza \\
    corona & vaxxer & Robert Mello & Michael Houghton \\
     & anti-vaxxer & Mario Capecchi  & Charles Rice \\
     & vaxer & Elizabeth Blackburn & Luc Montagnier \\ 
     & anti-vaxer & Jules Hoffmann & Jeffrey Hall \\
     & & Shinya Yamanaka & \\
     \hline
    \end{tabular}
    \label{tab:search}
\end{table}

Laureate names were enclosed in double quotations. OR statements were applied down columns, AND statements were applied across columns. The Twitter API is not case-sensitive and dashes also return condensed versions (for example, covid-19 also returns covid19, and covid also returns COVID, Covid, etc).

For the generic COVID dataset, the ``vaccine'' column was ejected. For the baseline dataset, both the ``covid'' and ``vaccine'' columns were ejected.

\end{document}